\documentclass[a4paper,10pt]{article}
\pdfoutput=1 %

\usepackage{jheppub} %

\usepackage[T1]{fontenc} %

\usepackage[pass]{geometry}
\newlength\DX \DX=3cm
\paperwidth=\dimexpr\paperwidth-\DX\relax
\hoffset=\dimexpr\hoffset-.5\DX\relax
\newlength\DY \DY=3cm
\paperheight=\dimexpr\paperheight-\DY\relax
\voffset=\dimexpr\voffset-.1\DY-.5\footskip\relax

\usepackage[utf8]{inputenc}
\usepackage[english]{babel}
\usepackage{hyperref}
\usepackage{amsmath}
\usepackage{amssymb}
\usepackage{amsfonts}
\usepackage{amsthm,bm}
\usepackage{physics}
\usepackage{aas_macros}
\usepackage{cleveref}
\usepackage{graphicx}
\usepackage{caption}
\usepackage{subcaption}

\newcommand{\sch}{Schwarzschild }

\title{Nonlinear Quasi-Normal Modes: Uniform Approximation}
\author{Bruno Bucciotti,}
\author{Adrien Kuntz,}
\author{Francesco Serra,}
\author{Enrico Trincherini}

\affiliation{Scuola Normale Superiore, Piazza dei Cavalieri 7, 56126, Pisa, Italy and INFN Sezione di Pisa, Largo Pontecorvo 3, 56127 Pisa, Italy}

\emailAdd{bruno.bucciotti@sns.it}
\emailAdd{adrien.kuntz@sns.it}
\emailAdd{francesco.serra@sns.it}
\emailAdd{enrico.trincherini@sns.it}

\abstract{Recent works have suggested that nonlinear (quadratic) effects in black hole perturbation theory may be important for describing a black hole ringdown. We show that the technique of uniform approximations can be used to accurately compute 1) nonlinear amplitudes at large distances in terms of the linear ones, 2) linear (and nonlinear) quasi-normal mode frequencies, 3) the wavefunction for both linear and nonlinear modes. Our method can be seen as a generalization of the WKB approximation, with the advantages of not losing accuracy at large overtone number and not requiring matching conditions. To illustrate the effectiveness of this method we consider a simplified source for the second-order Zerilli equation, which we use to numerically compute the amplitude of nonlinear modes for a range of values of the angular momentum number.
}

\begin{document}

\maketitle

\section{Introduction}

Quasi-normal modes (QNM) of black holes (BH) are a highly valuable tool for accurately describing the last part of a gravitational wave (GW) signal, known as the ringdown phase~\cite{Berti_2009}. Far from the BH, the metric perturbation is expressed as a sum of exponentially decaying oscillations. One remarkable result of BH perturbation theory is that the frequency of these oscillations is quantized, meaning that it depends on a integer <<overtone number>> $n$ and on the spherical harmonic numbers $(\ell, m)$~\cite{89c00a43-362f-3c19-9cc2-fa0808f91935}. While QNMs do not form a complete basis for describing the full late-time signal of a BH collision ~\cite{PhysRevD.34.384}, they still fit very well the results from numerical relativity~\cite{Buonanno:2006ui,Berti:2007dg}.

Various techniques have been developed to compute QNM frequencies with high precision. All of these techniques rely on studying the simplified problem of a monochromatic linearized perturbation on \sch or Kerr spacetime obeying boundary conditions infalling at the BH horizon and outgoing at infinity. 
To date, the most accurate of these  is Leaver's continued-fraction method~\cite{leaver}. Another well-known approximation technique is an application of WKB method to the QNM problem~\cite{PhysRevD.35.3621,Iyer:1986vv,1985ApJ...291L..33S,2003PhRvD..68b4018K,Matyjasek_2017}. This last method has the advantage of being analytic, thus providing insights into the mechanism of QNM generation and allowing generalizations to beyond-GR theories~\cite{Franciolini:2018uyq,Hui:2022vov}. Other techniques include the Bender-Wu approach~\cite{Hatsuda:2023geo}, the monodromy method~\cite{motl2003asymptotic} (see e.g.~\cite{Berti_2009} for a review), methods based on a spectral decomposition of the equations
in hyperboloidal slices~\cite{Ansorg:2016ztf, Ripley:2022ypi} and more recent developments related to supersymmetric and conformal field theories (see e.g.~\cite{Aminov:2020yma,Bonelli:2021uvf,Bonelli:2022ten,Aminov:2023jve}). Finally, closely related to the WKB method, the method of uniform expansions was developed in~\cite{Hatsuda:2021gtn}, where it was applied together with Borel-Padé summation.

In this article we will also rely on the method of uniform expansions (see e.g.~\cite{Berry_1972,marino_2021}), performing detailed computations of the linear frequencies for various $n$ and $\ell$ and extending it to compute the amplitude of nonlinear QNMs. We will provide a detailed explanation of this technique in~\cref{sec:uniform}. Interestingly, this method allows to circumvent the matching conditions traditionally used in the Schutz and Will WKB approach~\cite{1985ApJ...291L..33S}. Indeed, our approximate QNM wavefunction is accurate both near the maximum of the potential and at large distances from it. Similar to the WKB method, this approach does not strongly depend on the precise form of the potential in GR, making it potentially applicable to other modified-gravity settings. However, unlike the WKB method, we will demonstrate in~\cref{sec:linear} that its accuracy remains high even as the overtone number $n$ increases. We will achieve this without the need for high-order matching, resulting in more compact expressions compared to the WKB method.

Our primary motivation for introducing this new analytic technique is related to the study of nonlinearities in BH ringdown. Indeed, since the merger of two BHs is a highly nonlinear process, it is not surprising that the ringdown phase may not be entirely described by a linearized perturbation on the BH background. Several works have highlighted the significance of nonlinearities in BH ringdown~\cite{Ma_2022,London_2014,Mitman:2022qdl, Cheung:2022rbm,Nakano:2007cj,Ioka:2007ak,Lagos_2023}. In particular, due to mode-coupling effects at quadratic order in perturbation theory, there exist a set of <<nonlinear quadratic QNMs>> whose frequencies are given by the sum or difference of linear QNM frequencies~\cite{Lagos_2023}. These modes emerge at the second order in perturbation theory because the Regge-Wheeler and Zerilli equations involve a source term with a product of two linear QNMs~\cite{Brizuela:2006ne, Brizuela:2009qd, Brizuela:2007zza}. Notice, however, that the complete gravitational-wave signal after a BH merger does not only consist in a superposition of QNMs, as it also contains a flat-space and tail piece both at first and second-order~\cite{PhysRevD.55.468,PhysRevD.34.384,Lagos_2023,Okuzumi:2008ej,Carullo:2023tff}; our work only concentrates on the QNM part of the Green's function.

Including these nonlinearities into ringdown models could prove to be advantageous for fitting the signals.
Given the anticipated increase in precision of BH spectroscopy made possible by the space-based interferometer LISA~\cite{Berti_2006}, it is crucial to gain a quantitative understanding of the structure of these nonlinear QNMs, both analytically and numerically. In the following, we will loosely use the term "nonlinear" to explicitly refer to second-order perturbation theory of BH spacetimes, even if in full generality the nonlinearities also contain cubic and higher-order perturbations.

An open problem in the study of nonlinear QNMs, which this article aims to contribute to, is determining the amplitude of these modes. In ringdown models, the amplitude of linear QNMs is typically a free parameter dependent on initial conditions, which should be fitted against data~\cite{Berti_2006}. Being generated by nonlinear processes involving the multiplication of linear modes, we expect that the amplitude of nonlinear QNMs can be {\it entirely} determined by the amplitudes of the linear modes themselves. Such a statement has already been proved in numerous previous works~\cite{Lagos_2023,Okuzumi:2008ej,Cheung:2022rbm,Mitman:2022qdl,Nakano:2007cj,Ioka:2007ak,Kehagias:2023ctr,Redondo-Yuste:2023seq}. By considering an idealized model of a pure nonlinear QNM that satisfies the same boundary conditions as the linear ones, we will once again demonstrate that this intuition is true. In other words, our work aims to quantitatively answer the question: <<Given two linear QNMs with amplitudes $\mathcal{A}_1$ and $\mathcal{A}_2$, what is the amplitude $\mathcal{A}_\mathrm{NL}$ of the associated nonlinear QNM?>>. Knowing how to accurately solve this open problem would greatly benefit ringdown modeling, as it would make it possible to account for nonlinearities without introducing redundant free parameters at the level of data analysis. 

Our strategy concerning the study of nonlinear QNM is the following. First, we will introduce in~\cref{sec:nonlinDefs} the equation obeyed by second-order perturbations of the metric field around a \sch black hole, which turns out to be a Regge-Wheeler (RW)/Zerilli equation with a source term containing the nonlinearities~\cite{Brizuela:2006ne, Brizuela:2007zza, Ioka:2007ak, Nakano:2007cj, Brizuela:2009qd, Spiers:2023mor, Gleiser:1995gx, Gleiser_2000}. When dealing with the nonlinear source term, particular care should be given to the choice of gauge as its asymptotic behavior crucially depends on that choice~\cite{Gleiser:1995gx}. For simplicity and to focus the discussion on the analytic method we developed to solve the RW/Zerilli equation, we will use a simplified expression for the source term featuring the correct asymptotic behavior.  It's important to note that there are no obstructions in using the full source term, as we show in a concrete example in~\cref{app:source}.
In~\cref{sec:nonlin} we next employ the method of uniform expansions introduced in~\cref{sec:uniform} to obtain an analytical approximation to the solution of this second-order equation. This allows us to write in~\cref{eq:ratioNumerical} the amplitude of the nonlinear QNM at large distance, which we subsequently evaluate numerically in~\cref{table:nonlinear_ratio,table:nonlinear_ratio2}. We finally estimate the accuracy of the method of uniform expansion in~\cref{sec:accuracy}, finding that the approximate solutions should be correct at the percent level. 
\smallskip

Except for section~\ref{sec:uniform}, we will denote all linear quantities with an overbar ($\bar\Psi$) in order to distinguish them from nonlinear ones ($\Psi$).
Our convention for the Fourier transform is
\begin{equation} \label{eq:convFourier}
    \Psi(t) = \int \frac{\mathrm{d} \omega}{2 \pi} e^{i \omega t} \Psi(\omega)
\end{equation}
Newton's constant is $G$. All plots in this article are displayed in units where $GM=1$, where $M$ is the mass of the black hole.
\smallskip

\textbf{Note added:}
As we were finalizing the writing of this article, we became aware of a related work~\cite{perrone2023nonlinear}, similar to ours in spirit. While using a different technique than the one presented in~\cite{perrone2023nonlinear}, we can confirm their main conclusions concerning the study of nonlinear QNMs. For example, our estimates for the amplitudes of nonlinear QNMs agree in the particular case illustrated in~\cref{app:source}.
However, in addition to presenting new results concerning also the study of linear QNM frequencies, we believe that our contribution offers a more quantitative estimate of the amplitudes of nonlinear QNMs, as we evaluate the accuracy of our approximation, see ~\cref{sec:accuracy}. As an additional difference, we insist on requiring the appropriate asymptotic behavior for the term sourcing the nonlinearities (the \textquoteleft source\textquoteright{} term) in the second order RW/Zerilli equation (see~\cref{sec:nonlinDefs}). This requirement is necessary to ensure that second-order QNMs are well defined. This point appears to be less emphasized in Ref.~\cite{perrone2023nonlinear}. %

Furthermore, another recent work, Ref.~\cite{pan2023uniform}, employs uniform approximations to study the Poschl-Teller potential, without focusing on the QNMs, but rather studying the accuracy of the method.
\section{Nonlinear perturbations of a \sch black hole} \label{sec:nonlinDefs}

It is well-known that linear perturbations around a \sch black hole of mass $M$ obey the Regge-Wheeler~\cite{Regge:1957td} (RW) or Zerilli~\cite{Zerilli:1970aa} equation, depending on the parity of the mode. Focusing on eigenfunctions at definite frequency $\bar\omega$ and angular momentum $(\bar\ell,\bar m)$, and denoting by $\bar \Psi(r_*)$ the dimensionless amplitude of the propagating degree of freedom at linear order (either in the even or the odd sector), we have
\begin{equation}
    \frac{\mathrm{d}^2 \bar \Psi}{\mathrm{d}r_*^2} + \big( \bar\omega^2 - V(r) \big) \bar \Psi = 0 \; ,
\end{equation}
where $r_*$ is the tortoise coordinate $r_* = r + 2 G M \log \big( r/(2GM)-1 \big)$ (while $r$ is the standard \sch radial coordinate), and $ V(r)$ is the Regge-Wheeler or Zerilli potential, given by
\begin{align} \label{eq:RWZPot}
    V^\mathrm{RW}(r) &= \bigg( 1 - \frac{2 G M}{r} \bigg) \bigg( \frac{\ell(\ell+1)}{r^2} - \frac{6 G M}{r^3} \bigg) \; , \nonumber \\
    V^\mathrm{Z}(r) &= \bigg( 1 - \frac{2 G M}{r} \bigg) \frac{2 \lambda^2(\lambda+1)r^3 + 6 \lambda^2 G M r^2 + 18 \lambda G^2 M^2 r + 18 G^3 M^3}{r^3(\lambda r + 3 G M)^2} \; , \quad \lambda = \frac{(\ell-1)(\ell+2)}{2} \; .
\end{align}
At second order in perturbation theory, it turns out that the second-order perturbations of the metric can also be encoded into one parity-odd and one parity-even function which we will denote by $\Psi$ (also chosen to be dimensionless in the following)~\cite{Brizuela:2006ne, Brizuela:2007zza, Ioka:2007ak, Nakano:2007cj, Brizuela:2009qd, Spiers:2023mor, Gleiser:1995gx, Gleiser_2000}. These second-order amplitudes satisfy the same RW/Zerilli equation, albeit with a source term that is proportional to the product of linear modes: 
\begin{equation} \label{eq:RWZ}
    \frac{\mathrm{d}^2 \Psi}{\mathrm{d}r_*^2} + \big( \omega^2 - V(r) \big) \Psi = S (r) \; . %
\end{equation}
The explicit expression of $S(r)$ depends on the choice of gauge and dynamical variable $\bar\Psi$, see e.g.~\cite{Brizuela:2009qd, Nakano:2007cj} for some specific examples. Generically, the source consists in the product of two first-order modes $\bar \Psi_1$ and $\bar \Psi_2$ and their derivatives multiplied by functions of $r$, constructed so as to be quadratic in the linear amplitudes. Furthermore, even for a given parity of the second-order mode $\Psi$, the source can depend on \textit{both} even and odd-parity first-order modes.
Thus, we will rewrite $S$ as
\begin{equation} \label{eq:defSHat}
    S(r) = \hat S(r) \bar \Psi_1(r)  \bar \Psi_2(r) \; ,
\end{equation}
where now $\hat S$ does not scale with the linear amplitudes~\footnote{For example, $\hat S$ can contain the ratio $\bar \Psi_1' / \bar \Psi_1$ which does not scale with the overall linear amplitude.}. It can further be shown that, with an appropriate gauge choice, one can impose the following asymptotic behavior of $\hat S$~\cite{Gleiser:1995gx, Nakano:2007cj}:
\begin{align} \label{eq:asymptoticsShat}
    \hat S(r) &= \mathcal{O} \big( r^{-2} \big) \; , \quad \mathrm{for} \; r \rightarrow \infty \; , \nonumber \\
     \hat S(r) &= \mathcal{O} \bigg( 1 - \frac{2 G M}{r} \bigg) \; , \quad \mathrm{for} \; r \rightarrow 2 G M \; .
\end{align}
These asymptotics are crucial for the study of nonlinear quasi-normal modes. This requirement is not specifically linked to our method; any study of nonlinear QNMs should employ dynamic variables $\bar\Psi_{1,2}$ defined so that \cref{eq:asymptoticsShat} are ensured, otherwise second-order QNMs would display a divergent power-law scaling, as discussed in Ref.~\cite{Lagos_2023}. We will later observe that this translates in our formalism into the convergence of the integrals described in~\cref{app:CVint}.

In order to keep the discussion as simple as possible, in the spirit of Refs.~\cite{Okuzumi:2008ej,Lagos_2023} we will use the following toy-model for $\hat S$:
\begin{equation} \label{eq:toyModelS}
    \hat S(r) = \frac{1}{r^2} \bigg( 1 - \frac{2GM}{r} \bigg) \; ,
\end{equation}
which respects the boundary conditions~(\ref{eq:asymptoticsShat}). This choice is mainly motivated by two considerations:
\begin{enumerate}
    \item Our focus in this article is demonstrating the effectiveness of an analytic method for computing quasi-normal mode amplitudes at the second order, rather than reconstructing a complete waveform template accounting for second-order QNMs.
    As a result, we do not (at this point) aim at a maximally realistic source term. When using our results to estimate the quantitative impact of second-order modes on ringdown waveforms, it will be important to go beyond the toy-model approximation (\cref{eq:toyModelS}) and use the complete expression of $S$ in a gauge adapted for gravitational radiation. We leave this issue to further work.
    \item Expressions for the source $S$ featuring the correct asymptotic behavior are rare in the literature. To our knowledge, ready-to-use expressions for $S$ are only available for specific values of $\ell$ in Refs.~\cite{Nakano:2007cj, Brizuela:2009qd, Gleiser:1995gx}. In addition, the only one that exhibits the asymptotics~(\ref{eq:asymptoticsShat}) (particularly near the horizon) is the source present in~\cite{Nakano:2007cj}, which however focuses only on $(\ell=2)\times(\ell=2)\rightarrow(\ell=4)$.  We further show in~\cref{app:source} that our toy-model (or simple modifications thereof) can quite accurately fit the function $\hat S$ provided in~\cite{Nakano:2007cj}.
\end{enumerate}

Having presented the necessary details of the dynamics we are interested in, we now turn to introduce our method for solving the RW/Zerilli equation.

\section{ Uniform expansions} \label{sec:uniform}

In this section, we introduce the concept of uniform expansions, first set out in~\cite{PhysRev.91.174, dingle1956method}. We refer the reader to~\cite{Berry_1972,marino_2021} for more details and applications of the formalism to quantum-mechanical problems.
The basic idea is to compare the Zerilli/RW equation~\eqref{eq:RWZ} with a simpler, \emph{auxiliary} differential equation which is exactly solvable. While this step sounds similar to what we would do with a Poschl-Teller potential, we will then relate the exact solutions of the new auxiliary differential equation to \emph{approximate} solutions of the original one.
We will freely switch between $r$ and $r_*$ inside the argument of functions, leaving the conversion $r_*(r)$ implicit. In this section we will not distinguish between $\Psi$ and $\bar\Psi$, since the discussion applies to both.
\smallskip

We choose to consider the following auxiliary differential equation for $\phi(\sigma)$, where $\sigma(r)$ will be interpreted as a local rescaling of $r$
\begin{equation} \label{eq:DEphi}
    \frac{\mathrm{d}^2 \phi}{\mathrm{d} \sigma^2} + \Theta (\sigma) \phi = \Lambda (\sigma) \; ,
\end{equation}
where $\sigma(r)$ and $\Lambda(\sigma)$ are two functions we will solve for later on. We now want to choose the potential $\Theta$ to be \textquotedblleft close\textquotedblright\, in some sense to $\omega^2 - V(r)$ in Eq.~\eqref{eq:RWZ}. We know that physically quasi-normal modes are generated near the maximum of the potential present in Eq.~\eqref{eq:RWZ}, so that this region should matter the most. In order to exploit this, we choose $\Theta$ to be a quadratic function of $\sigma$
\begin{equation}\label{eq:defGamma}
    \Theta(\sigma) = i \bigg( \nu + \frac{1}{2} \bigg) + \frac{\sigma^2}{4} \; ,
\end{equation}
where we have introduced a parameter $\nu$ and some convenient normalizations. The two solutions of the homogeneous part of equation~\eqref{eq:DEphi} are
\begin{equation}\label{eq:phi1phi2}
    \phi_A(\sigma) = D_\nu \big( e^{i \pi/4} \sigma \big) \; , \quad  \phi_B(\sigma) = D_{-\nu -1} \big( e^{3i \pi/4} \sigma \big) \; .
\end{equation}
where $D_\nu$ is the parabolic cylinder function.
We now relate the exact solutions $\phi_A,\phi_B(\sigma)$ to approximate solutions of equation~(\ref{eq:RWZ}), thanks to a local rescaling that will transform $r$ to $\sigma$. The function $\sigma(r)$ will now be determined by simply substituing the following ansatz
\begin{equation}
    \Psi(r) = f(r) \phi \big( \sigma(r) \big) \; ,
\end{equation}
into Eq.~\eqref{eq:RWZ}. $f(r)$ will later be chosen in a convenient way. Plugging into equation~\eqref{eq:RWZ} and using the auxiliary differential equation~\eqref{eq:DEphi} on $\phi''$ we obtain
\begin{equation}
    0 = \big( f'' + \big( \omega^2 - V \big) f - (\sigma')^2 f \, \Theta \big) \phi + \big( 2 f' \sigma'+ f \sigma'' \big) \frac{\mathrm{d} \phi  }{\mathrm{d}\sigma} + (\sigma')^2 f \, \Lambda - S \; ,
\end{equation}
where a prime denotes differentiation with respect to the tortoise coordinate $r_*$. We now choose the following expressions for $f$ and $\Lambda$:
\begin{equation} \label{eq:defLambda}
    f = (\sigma')^{-1/2} \; , \quad \Lambda = \frac{S}{(\sigma')^{3/2}} \; .
\end{equation}
This choice makes everything outside the first parenthesis vanish. We are thus left with a differential equation on the only remaining undetermined variable $\sigma$, because $\phi$ factors out:
\begin{equation} \label{eq:fullEqSigma}
    \omega^2 - V = (\sigma')^2 \, \Theta - (\sigma')^{1/2} \frac{\mathrm{d}^2}{\mathrm{d}r_*^2} (\sigma')^{-1/2} \; .
\end{equation}
The last term is called the schwarzian derivative (up to a normalization).
If $\Theta(\sigma)$ has been chosen appropriately, $\sigma$ will be a slowly varying function of $r_*$ and the schwarzian should be negligible. It is worth pausing to mention that the usual WKB method would have been recovered had we chosen $\Theta(\sigma)\equiv\pm1$~\cite{Berry_1972}; the schwarzian is the parameter controlling the validity of the approximation for that case also. This should reassure the reader that, despite our crude auxiliary potential $\Theta(\sigma)$, great accuracy can be achieved with this method~\footnote{A uniformly valid Airy matching is also easily obtained by considering $\Theta(\sigma)=\pm\sigma$~\cite{Berry_1972}}. To achieve even better results, one could perturbatively include the effects of the schwarzian derivative and Borel-Padé resum the resulting series as suggested in~\cite{Hatsuda:2021gtn}.

Assuming for now that the schwarzian is negligible, we can approximate
\begin{equation} \label{eq:dsigmadr}
    \frac{\mathrm{d}\sigma}{\mathrm{d}r_*} \simeq \bigg( \frac{\omega^2 - V(r)}{\Theta(\sigma(r))} \bigg)^{1/2} \; ,
\end{equation}
where the sign choice matches $\sigma$ to $r_*$ preserving orientation. Notice that this is the only approximation we have made so far. It is possible to go beyond this approximation and include higher-order corrections, see e.g.~\cite{doi:10.1143/JPSJ.14.1771,1971JChPh..54.3864P,PhysRev.106.1156}, although we will not do it here. Eq.~\eqref{eq:dsigmadr} completely determines $\sigma$ up to an additive shift.
\smallskip

WKB approximation is invalid near to the turning points, that is the zeros of $\omega^2-V$. This is readily seen in this framework, because the schwarzian derivative could blow up. To ensure it remains finite in the whole domain of interest (i.e. the real $r_*$ line), turning points should be matched so that $\frac{\dd \sigma}{\dd r_*}$ remains finite; the schwarzian derivative is then not necessarily divergent (necessary condition for the approximation to work). This will impose a relation between $\omega$ and $\nu$. The (uniform) smallness of the schwarzian compared to $\omega^2-V$ (sufficient condition) is to be assessed by hand (see~\cref{sec:accuracy} and also~\cite{Berry_1972} for a discussion).

Let us denote by $\sigma^-$ and $\sigma^+$ the two zeros of $\Theta$, and by $r_*^-$ and $r_*^+$ the equivalent zeros of $\omega^2-V$ i.e. its turning points.
First, we fix the shift ambiguity of $\sigma$ by setting $\sigma(r_*^+)=\sigma^+$. This choice has both the numerator and the denominator inside the square root of Eq.~(\ref{eq:dsigmadr}) vanish linearly, keeping the ratio finite. Then integrating equation~(\ref{eq:dsigmadr}) gives

\begin{equation}\label{eq:defsigma}
     \int_{\sigma^+}^{\sigma} \Theta^{1/2}(\tilde \sigma) \mathrm{d} \tilde \sigma = \bigg[ \frac{\sigma}{2} \Theta^{1/2} + i \bigg( \nu+\frac{1}{2} \bigg) \log \big( \sigma + 2 \Theta^{1/2} \big) \bigg]_{\sigma^+}^{\sigma} = \int_{r_*^+}^{r_*} \big( \omega^2 - V \big)^{1/2} \mathrm{d} \tilde r_* \; ,
\end{equation}
where we have used the explicit form of $\Theta$ in Eq.~\eqref{eq:defGamma}.
This implicitly defines the coordinate $\sigma$. In addition, for the schwarzian in Eq.~(\ref{eq:fullEqSigma}) to remain bounded,   by integrating Eq.~\eqref{eq:dsigmadr} between the turning points we get an \textquotedblleft area law\textquotedblright
\begin{equation}
    \int_{\sigma^-}^{\sigma^+} \Theta^{1/2} \mathrm{d} \sigma = \int_{r_*^-}^{r_*^+} \big( \omega^2 - V(r) \big)^{1/2} \mathrm{d}r_* \; .
\end{equation}
Using $\sigma_\pm = \pm (1 - i) \sqrt{2 \nu +1}$, we get a constraint for the integral of the potential near the turning points:
\begin{equation} \label{eq:arealaw}
    \int_{r_*^-}^{r_*^+} \big( \omega^2 - V(r) \big)^{1/2} \mathrm{d}r_* = \pi \bigg( \nu + \frac{1}{2} \bigg) \; .
\end{equation}
This condition will be used to obtain the value of $\omega$ when solving for linear quasi-normal modes (where we will impose a quantization condition on $\nu$). Alternatively, it will give the value of $\nu$ for nonlinear quasi-normal modes, where this time we will start from a given value of $\omega$.

To recap, we now have found the generic solution of Eq.~\eqref{eq:RWZ}:
\begin{equation} \label{eq:Psisol}
    \Psi(r) = \bigg( \frac{\Theta(\sigma(r))}{\omega^2 - V(r)} \bigg)^{1/4} \phi \big( \sigma(r) \big) \; ,\quad \text{(Uniform approximation)}
\end{equation}
where $\phi$ is a solution to the simpler equation~\eqref{eq:DEphi}, $\Theta$ is given in Eq.~\eqref{eq:defGamma} and $\sigma$ is determined from Eq~\eqref{eq:defsigma}. We highlight that, contrary to WKB approximate solutions, Eq.~(\ref{eq:Psisol}) exhibits no divergence close to the turning points we matched, making our expression ready to plot (see~\cref{fig:psi}). We now have to impose the correct boundary conditions to this solution. Let us discuss separately the cases of linear and nonlinear modes.

\section{Linear modes}\label{sec:linear}

We study linear modes of angular number $\bar \ell$ which will be different from the nonlinear angular number $\ell$. The linear mode amplitude $\bar \Psi$ obeys Eq.~\eqref{eq:RWZ} where the source term $S$ is set to zero. The auxiliary field $\bar \phi$ obeys the homogeneous part of equation~\eqref{eq:DEphi}, whose solutions are given in Eq.~\eqref{eq:phi1phi2}. From Eq.~\eqref{eq:Psisol} this means that
\begin{equation} \label{eq:barpsiAB}
    \bar \Psi(r) = \bigg( \frac{\bar \Theta(\bar \sigma )}{\bar \omega^2 - \bar V(r)} \bigg)^{1/4} \bigg( \mathcal{A} \, D_{\bar\nu} \big( e^{i \pi/4} \bar \sigma \big) + \mathcal{B} \, D_{-\bar\nu-1} \big( e^{3i \pi/4} \bar \sigma \big) \bigg) \; ,
\end{equation}
where $\mathcal{A}$ and $\mathcal{B}$ are two amplitudes that are determined by initial conditions and boundary conditions at the horizon and infinity, and we have introduced an overbar on all quantities related to first-order modes. Notice that, in order to make the field $\bar \Psi$ dimensionless, the amplitudes should scale as $\mathcal{A}, \mathcal{B} \propto \bar \omega^{1/2}$. Additionally, the fundamental assumption that $\bar \Psi$ is a small perturbation of a background \sch spacetime translates into the fact that its amplitude is small, i.e. $\bar \omega^{-1/2} \mathcal{A} \ll 1$, $\bar \omega^{-1/2} \mathcal{B} \ll 1$. This assumption will ensure that second-order modes are always smaller than first-order (linear) ones, as they should be proportional to the square of the linear amplitude, see~\cref{eq:defSHat}.
Now, it remains to impose the correct boundary conditions. To ensure that waves leave the system at infinity and enter the black hole close to its horizon, we must impose the following QNM boundary conditions:

\begin{align}\label{eq:QNMBC1}
    \bar \Psi(r) &\sim \bar \Psi_\infty e^{-i \bar \omega r_*} \; , \quad \mathrm{for} \; r_* \rightarrow \infty \; ,\\
    \bar \Psi(r) &\sim \bar \Psi_H e^{i \bar \omega r_*} \; , \quad \mathrm{for} \; r_* \rightarrow - \infty \label{eq:QNMBC2} \; ,
\end{align}
where the choice of sign inside of the exponential is dictated from our convention for the Fourier transform~\eqref{eq:convFourier}, and $\bar \Psi_\infty$ and $\bar \Psi_H$ are two constants. The following subsection works out the constraints that Eq.~(\ref{eq:QNMBC1}) imposes on $\mathcal{A},\mathcal{B}$.

\subsection{Quantization condition}
We now determine the quantization condition that will give the quasi-normal modes spectrum. At the same time, we compute the asymptotic behavior of the wavefunction at large $|r_*|$ accurately to $\mathcal{O}(1)$ phases, because it will play a role in section~\ref{sec:nonlin}. The hasty reader that is only interested in the quantization condition should focus on equations~(\ref{eq:barsigmainfty}, \ref{eq:barsigmaminfty}, \ref{eq:Dnuinfty}-\ref{eq:Dmnuminfty}) and equations~(\ref{eq:QNMBC1}, \ref{eq:QNMBC2}), jumping straight to \cref{eq:quantization_for_Theta}.
\smallskip

In order to determine the behavior of $\bar \Psi$ in Eq.~\eqref{eq:barpsiAB} we have to relate the asymptotic behavior of $\bar \sigma$ and $r_*$. By noticing that $V$ vanishes and its integral is finite for both large positive and negative $r_*$, we get from Eq.~\eqref{eq:defsigma} the following asymptotic behavior:
\begin{align} \label{eq:barsigmainfty}
    \bar \sigma &= 2 \big( \bar \omega r_* \big)^{1/2} + \mathcal{O} \bigg( \frac{\log r_*}{r_*^{1/2}} \bigg) \; , \quad \quad \mathrm{for} \; r_* \rightarrow \infty \; , \\
    \bar \sigma &= - 2 \big( - \bar \omega r_* \big)^{1/2} + \mathcal{O} \bigg( \frac{\log (- r_*)}{(-r_*)^{1/2}} \bigg) \; , \quad \quad \mathrm{for} \; r_* \rightarrow - \infty \; . \label{eq:barsigmaminfty}
\end{align}
We will need in the following the expression of $\bar \sigma^2$ up to $\mathcal{O}(1)$. Evaluating the expansion of Eq.~\eqref{eq:defsigma} to the next order, we get
\begin{align} \label{eq:barsigma2infty}
    \frac{\bar \sigma^2}{4} = \bar \omega r_* + \bar C_{\infty} - \frac{i}{2} \bigg( \bar \nu + \frac{1}{2} \bigg) \log (4 \bar \omega r_*) + \mathcal{O} \bigg( \frac{\log r_*}{r_*} \bigg) \; , \quad \quad \mathrm{for} \; r_* \rightarrow \infty \; , \\
    \frac{\bar \sigma^2}{4} = - \bar \omega r_* + \bar C_{H} - \frac{i}{2} \bigg( \bar \nu + \frac{1}{2} \bigg) \log (- 4 \bar \omega r_*) + \mathcal{O} \bigg( \frac{\log (- r_*)}{r_*} \bigg) \; , \quad \quad \mathrm{for} \; r_* \rightarrow - \infty \; , \label{eq:barsigma2minfty}
\end{align}
where $\bar C_\infty$ and $\bar C_H$ are two constants given by
\begin{align} \label{eq:barCinfty}
    \bar C_\infty = - \bar \omega \bar r_*^+ + \int_{\bar r_*^+}^\infty \bigg[ \big(\bar  \omega^2 - \bar V \big)^{1/2} - \bar \omega \bigg] \mathrm{d} \tilde r_*  + \bigg(\bar \nu + \frac{1}{2} \bigg)  \bigg[ \frac{\pi}{4} + \frac{i}{2} \bigg( \log \big( 2\bar \nu+1 \big) -  \log 2 -1 \bigg) \bigg] \; , \\ \label{eq:barCH}
    \bar C_H =  \bar \omega \bar r_*^+ + \int_{-\infty}^{\bar r_*^+} \bigg[ \big(\bar  \omega^2 - \bar V \big)^{1/2} - \bar \omega \bigg] \mathrm{d} \tilde r_*  + \bigg( \bar \nu + \frac{1}{2} \bigg)  \bigg[ - \frac{3\pi}{4} + \frac{i}{2} \bigg( \log \big( 2\bar \nu+1 \big) - \log 2 -1 \bigg)  \bigg] \; .
\end{align}
Finally, there remains to obtain the behavior of the parabolic cylinder functions for large $\bar \sigma$, which are
\begin{align} \label{eq:Dnuinfty}
    D_{\bar \nu} \big( e^{i \pi/4} \bar \sigma \big) &\sim e^{i \pi \bar \nu /4} \bar \sigma^{\bar \nu} e^{-i \bar \sigma^2/4} \; , \\
    D_{-\bar\nu -1} \big( e^{3 i \pi/4} \bar \sigma \big) &\sim e^{- i \pi \bar\nu /4} \frac{\sqrt{2 \pi}}{\Gamma[1+\bar \nu]} \bar \sigma^{\bar\nu} e^{-i \bar \sigma^2/4} + e^{- 3 i \pi (1+\bar \nu) /4} \bar \sigma^{-\bar \nu -1} e^{i \bar \sigma^2/4} \; , \label{eq:Dmnuinfty}
\end{align}
for $\bar \sigma \rightarrow \infty$, and
\begin{align} \label{eq:Dnuminfty}
    D_{\bar \nu} \big( e^{i \pi/4} \bar \sigma \big) &\sim e^{-3 i \pi \bar \nu /4} (-\bar \sigma)^{\bar \nu} e^{-i \bar \sigma^2/4} +  e^{- i \pi (1+\bar \nu) /4} \frac{\sqrt{2 \pi}}{\Gamma[-\bar \nu]} (-\bar \sigma)^{-\bar \nu-1} e^{i \bar \sigma^2/4} \; , \\
    D_{-\bar \nu -1} \big( e^{3 i \pi/4} \bar \sigma \big) &\sim  e^{ i \pi (1+\bar \nu) /4} (-\bar \sigma)^{-\bar \nu -1} e^{i \bar \sigma^2/4} \; , \label{eq:Dmnuminfty}
\end{align}
for $\bar \sigma \rightarrow - \infty$. Plugging together \cref{eq:barpsiAB,eq:barsigmainfty,eq:barsigmaminfty,eq:barsigma2infty,eq:barsigma2minfty,eq:Dnuinfty,eq:Dmnuinfty,eq:Dnuminfty,eq:Dmnuminfty} we get to the following asymptotic behavior of $\bar \Psi$
\begin{align}
    \bar \Psi &\sim \frac{1}{\sqrt{2 \bar \omega}} \bigg[ e^{-i \bar \omega r_* -i \bar C_\infty} \bigg( \mathcal{A} e^{i \pi \bar \nu/4} + \mathcal{B} \frac{\sqrt{2\pi}}{\Gamma(1+\bar \nu)} e^{- i \pi \bar \nu/4} \bigg) + \mathcal{B} \, e^{i \bar \omega r_* +i \bar C_\infty} e^{- 3i \pi (1+ \bar \nu)/4} \bigg] \; , \; \mathrm{for} \; r_* \rightarrow \infty \\
    \bar \Psi &\sim \frac{1}{\sqrt{2 \bar \omega}} \bigg[ e^{-i \bar \omega r_* +i \bar C_H} \bigg( \mathcal{B} e^{i \pi (1+\bar \nu)/4} + \mathcal{A} \frac{\sqrt{2\pi}}{\Gamma(-\bar \nu)} e^{- i \pi (1+\bar \nu)/4} \bigg) + \mathcal{A} \, e^{i \bar \omega r_* -i \bar C_H} e^{- 3i \pi \bar \nu/4} \bigg] \; , \; \mathrm{for} \; r_* \rightarrow - \infty
\end{align}
We see that, in order to impose the QNM boundary conditions \cref{eq:QNMBC1,eq:QNMBC2} we have to choose
\begin{equation}
    \label{eq:quantization_for_Theta}
    \mathcal{B} = 0 \; , \quad \bar \nu = n \in \mathbb{N} \; ,
\end{equation}
since $1/\Gamma(-n) = 0$.

\subsection{Discussion}

\begin{figure}
    \centering
    \includegraphics[width=0.75\linewidth]{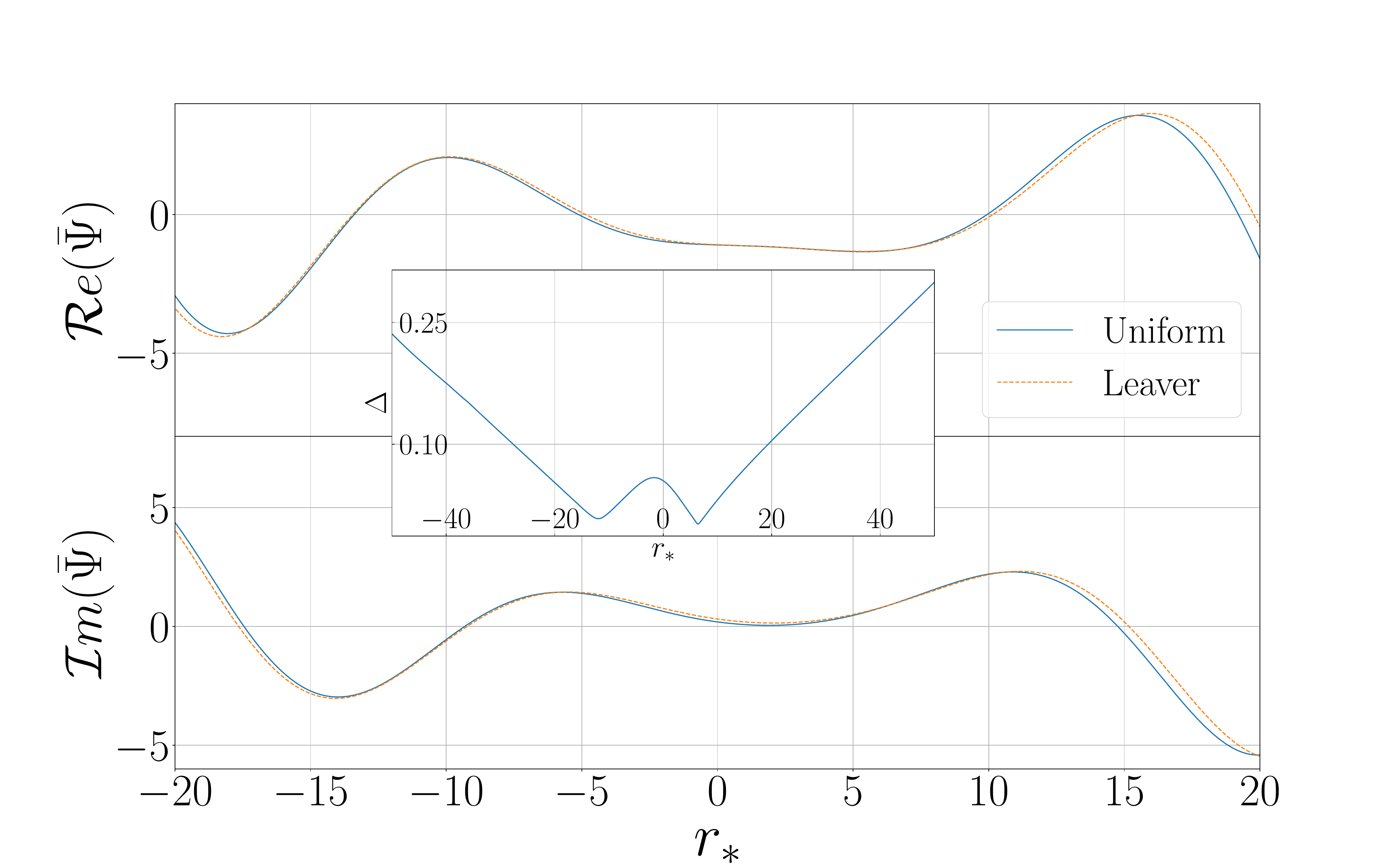}
    \caption{\textit{Wavefunction in the uniform approximation}: Plot of the real and imaginary parts of the wavefunction defined in~\cref{eq:barpsi} for $\bar \ell=2$, $n=0$ versus a very accurate numerical solution $\bar \Psi_\mathrm{Leaver}$ using Leaver's algorithm~\cite{leaver}, as a function of the tortoise coordinate $r_*$ and in units in which $GM=1$. The solutions are normalized so that $\bar \Psi \simeq e^{-i \bar \omega r_*}$ for $r_* \rightarrow \infty$. The inset shows the fractional deviation $\Delta = | (\bar \Psi - \bar \Psi_\mathrm{Leaver})/(\bar \Psi + \bar \Psi_\mathrm{Leaver}) |$ of the uniform approximation solution versus Leaver's wavefunction.  Our approximation is accurate both close to the maximum of $V$ and far from it, although the accuracy degrades for large values of $r_*$. %
    }
    \label{fig:psi}
\end{figure}

To recap, we have now found the (linear) solution to the homogeneous equation~\eqref{eq:RWZ} which is
\begin{equation} \label{eq:barpsi}
     \bar \Psi(r) =  \mathcal{A} \, \bigg( \frac{\bar \Theta(\bar \sigma )}{\bar \omega^2 - \bar V(r)} \bigg)^{1/4} \, D_{n} \big( e^{i \pi/4} \bar \sigma \big) \; ,
\end{equation}
where $\mathcal{A}$ is a small but otherwise arbitrary amplitude that can be fixed e.g. by the initial conditions of a ringdown signal. This solution is plotted in~\cref{fig:psi} for $\bar \ell=2$ and $n=0$, showing that as advertised in~\cref{sec:uniform} the profile for $\bar \Psi$ is accurate both close to the minimum of $V$ and at infinity.
The quantization condition giving the value of $\bar \omega_n$ is obtained from \cref{eq:arealaw}:
\begin{equation} \label{eq:quantization}
    \int_{\bar r_*^-}^{\bar r_*^+} \big(\bar  \omega_n^2 - \bar V(r) \big)^{1/2} \mathrm{d}r_* = \pi \bigg( n + \frac{1}{2} \bigg) \; .
\end{equation}
This equation can be numerically solved for $\bar \omega$ once given the RW/Zerilli potential in~\cref{eq:RWZPot}. On the technical side, we have to ensure that in \cref{eq:quantization} the integration path is chosen correctly.

\begin{figure}
    \centering
    \includegraphics[width=0.5\linewidth]{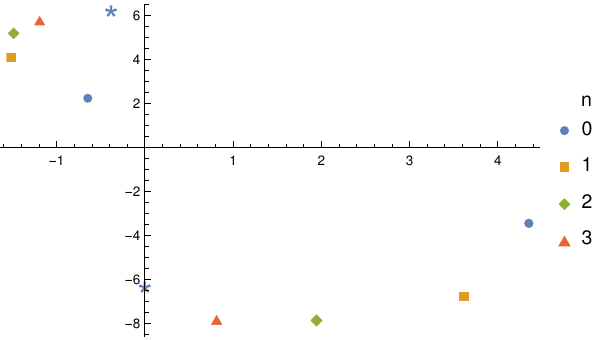}
    \caption{Turning points (in pairs) $\bar r_*^\pm$ for the Zerilli potential with $\bar \ell=2$ as a function of the overtone number $n$ in the complex plane, in units where $GM=1$. When $n$ increases, the turning points approach the two poles of the Zerilli potential (cyan $*$). This feature allows the integral in \cref{eq:quantization} to blow up as $n\rightarrow\infty$.}
    \label{fig:turning_points}
\end{figure}

This is nontrivial due to the presence of branch cuts and many Riemann sheets. The prescription we use is to start from real values of $\bar\omega$, where turning points $\bar r_*^\pm$ are real and the correct path is obviously identified as a straight line. As the imaginary part of $\bar \omega$ is increased, we track the turning points as they move away from the real axis. This prescription also correctly identifies which two of the many (complex) turning points of $\bar \omega^2-\bar V$ are to be used. Our integral in \cref{eq:quantization} is then completely well defined.
We plot the position of the turning points for $\bar \ell=2$ as a function of the overtone number in \cref{fig:turning_points}.
\smallskip

Alternatively, one can also recover the usual WKB formula of Schutz and Will~\cite{1985ApJ...291L..33S} by approximating the RW/Zerilli potential with a Taylor expansion up to second order near its maximum situated at $\hat r_*$:
\begin{equation} \label{eq:WKBquantization}
    \bar \omega^2 - \bar V(r) \simeq \bar \omega^2 - \bar V_0 - \frac{ \bar V_0''}{2} \big( r_* - \hat r_* \big)^2 \Rightarrow \frac{\bar \omega^2 - \bar V_0}{\sqrt{- 2 \bar V_0''}} = i \bigg( n + \frac{1}{2} \bigg) \; .
\end{equation}
The QNM frequencies given by our approximation \cref{eq:quantization}, and 6-th order WKB results are compared to the most accurate numerical estimates in \cref{table:linear_frequencies}.

The comparison reveals that 6-th order WKB performs better than our method for small $n$, albeit at the price of significant complexity: this comes as no surprise since we are only performing a lowest-order computation, and our result could be improved by taking into account the schwarzian in~\cref{eq:fullEqSigma}. Both approaches show improved accuracy as $\bar \ell$ increases. A remarkable feature of our approach is that, while WKB's agreement with the numerical solution worsens as $n$ increases, our approximation remains accurate independently of $n$; this is clearly showcased by looking at the (very extreme) $\bar \ell=2, n=1000$ mode.

We will now explain this behavior. Employing higher and higher order matching polynomials when using WKB has two positive, but distinct, effects
\begin{enumerate}
    \item It improves the approximation of $V$ that one uses very close to the top of the potential, which is crucial to obtain very precise results.
    \item $V$ is well approximated in a larger and larger domain, meaning that even for turning points which are quite separated (this happens for large $n$), the approximate potential is accurate.
\end{enumerate}
We already showed in \cref{eq:WKBquantization} that our method reconstructs a quadratic WKB matching. This is not surprising, since we used a quadratic auxiliary potential $\Theta(\sigma)$. The improvement that our method brings is that, thanks to the local deformations that $\sigma(r)$ accounts for, $V$ is now well approximated by $\Theta$ even \emph{far} from the maximum. Then, in some heuristic sense, our approximation works at first order in the first effect, but at all orders in the second one. Since the second effect determines how accuracy behaves with $n$, while the first one has more to do with \textquotedblleft absolute\textquotedblright{} accuracy, the results in \cref{table:linear_frequencies} can be understood. These observations also suggest that, if one were to increase the order of the polynomial $\Theta(\sigma)$ beyond $\sigma^2$, one would consistently do better than WKB at the same order.

A numerical fit of the relative errors of uniform approximations and 6-th order WKB using the data in \cref{table:linear_frequencies} indicates that uniform approximations should outperform 6-th order WKB roughly for
\begin{equation}
    n(\ell) \gtrsim 4+\ell/3 \; .
\end{equation}

Note that we have used the Zerilli potential for our numerical estimates, however since the Regge-Wheeler and Zerilli potential are isospectral~\cite{89c00a43-362f-3c19-9cc2-fa0808f91935} we could have as well used the RW potential. We preferred the Zerilli potential because for the RW potential in the complex $\bar r_*$ plane, the relevant turning points of $\bar\omega^2-\bar V^{\rm RW}$ approach the same limiting $\bar r_*$ point for large $n$, making the integral more susceptible to numerical instabilities. As can be seen from figure~\ref{fig:turning_points}, this is not the case for the Zerilli potential.

We now briefly explain how we computed the first column of~\cref{table:linear_frequencies}.

\paragraph{Efficient computation of $\bar\omega$}
Inverting equation~(\ref{eq:quantization}) to find $\bar \omega$ for given $n$ may seem like a daunting task. While (unless approached in special limits) the integral is not doable analytically, we mention how we obtained the frequencies without a painful scanning in $\bar \omega\in\mathbb{C}$.

If we have a target $\bar \nu\equiv n\in\mathbb{N}$ and a rough guess $\bar \omega_0$ of the related frequency, we can do the integral and compute its $\bar \nu_0$. Then the following holds (up to subleading corrections in $\delta\bar \omega$)
\begin{equation}
    \begin{split}
        \int (\bar \omega^2-\bar V)^{1/2}\dd r_* \simeq \int (\bar \omega_0^2-\bar V)^{1/2}\dd r_*+\bar \omega_0\,\delta\bar \omega\int\frac{1}{(\bar \omega_0^2-\bar V)^{1/2}}\dd r_* +\mathcal{O}(\delta\bar \omega^2)=\\
        =\pi\left(\bar \nu+\frac{1}{2}\right) = \pi\left(\bar \nu_0+\frac{1}{2}\right)+\pi\delta\bar \nu\,,\quad \delta\bar \nu\equiv\bar \nu-\bar \nu_0
    \end{split}
\end{equation}
where all integrals are all convergent and between the same turning points $\bar r_*^\pm$~\footnote{Note that, when expanding in $\delta\bar \omega$, the contribution coming from varying the extrema of the integral vanishes.}. We are then led to
\begin{equation}
    \delta\bar \omega = 
    \frac{\pi}{\bar \omega_0}(\bar \nu-\bar \nu_0) \left(\int\frac{1}{(\bar \omega_0^2-\bar V)^{1/2}}\dd r_*\right)^{-1}+\mathcal{O}(\bar \nu-\bar \nu_0)^2
\end{equation}
so that our guess $\bar \omega_0$ can be improved to $\bar \omega_0+\delta\bar \omega$ if $\delta\bar \nu$ is small. To obtain a guess, we used the frequency of modes with slightly smaller $n$ or $\bar \ell$; we then run the above algorithm, observing very fast convergence (1-2 iterations are sufficient). For $n=1000$ in \cref{table:linear_frequencies}, we used the numerical result as seed. A prediction independent of this input would require a more careful study of the convergence properties of our algorithm.

\begin{table}[ht!]
\centering
\begin{tabular}{cc|c|c|c}
	\hline
	$\bar \ell$ & $n$ & Uniform & 6-th order WKB & Exact \\
	\hline\hline
	2 & $0$ & $0.3854 + 0.0909 i$ ($3.1$\%) & $0.37371 + 0.08892 i$ ($0.014$\%) & $0.37367 + 0.08896 i$ \\
    \hline
    & $1$ & $0.3590 + 0.2796 i$ ($3.1$\%) & $0.34672 + 0.27388 i$ ($0.0089$\%) & $0.34671 + 0.27391 i$ \\
    \hline
    & $2$ & $0.3146 + 0.4868 i$ ($2.8$\%) & $0.30005 + 0.47883 i$ ($0.2$\%) & $0.30105 + 0.47828 i$ \\
    \hline
    & $3$ & $0.2670 + 0.7146 i$ ($2.4$\%) & $0.24551 + 0.71159 i$ ($1.2$\%) & $0.25150 + 0.70515 i$ \\
    \hline
    & $1000$ & $0.000 + 249.771 i$ ($0.06$\%) & $-$ & $0.051 + 249.618 i$ \\
    \hline\hline
    3 & $0$ & $0.6075 + 0.0935 i$ ($1.3$\%) & $0.59944 + 0.09270 i$ ($0.000049$\%) & $0.59944 + 0.09270 i$ \\
    \hline
    & $1$ & $0.5909 + 0.2837 i$ ($1.3$\%) & $0.58264 + 0.28129 i$ ($0.00088$\%) & $0.58264 + 0.28130 i$ \\
    \hline
    & $2$ & $0.5605 + 0.4830 i$ ($1.3$\%) & $0.55160 + 0.47906 i$ ($0.013$\%) & $0.55168 + 0.47909 i$ \\
    \hline
    & $3$ & $0.5215 + 0.6956 i$ ($1.3$\%) & $0.51111 + 0.69049 i$ ($0.1$\%) & $0.51196 + 0.69034 i$ \\
    \hline
    & $4$ & $0.4807 + 0.9219 i$ ($1.2$\%) & $0.46688 + 0.91799 i$ ($0.39$\%) & $0.47017 + 0.91565 i$ \\
    \hline
    & $5$ & $0.4428 + 1.1591 i$ ($1.1$\%) & $0.42437 + 1.16253 i$ ($1.$\%) & $0.43139 + 1.15215 i$ \\
    \hline\hline
    4 & $0$ & $0.8153 + 0.0946 i$ ($0.76$\%) & $0.80918 + 0.09416 i$ ($0.000014$\%) & $0.80918 + 0.09416 i$ \\
    \hline
    & $1$ & $0.8029 + 0.2857 i$ ($0.76$\%) & $0.79663 + 0.28433 i$ ($0.000038$\%) & $0.79663 + 0.28433 i$ \\
    \hline
    & $2$ & $0.7792 + 0.4822 i$ ($0.76$\%) & $0.77270 + 0.47990 i$ ($0.0018$\%) & $0.77271 + 0.47991 i$ \\
    \hline
    & $3$ & $0.7467 + 0.6870 i$ ($0.75$\%) & $0.73967 + 0.68390 i$ ($0.017$\%) & $0.73984 + 0.68392 i$ \\
    \hline
    & $4$ & $0.7088 + 0.9021 i$ ($0.73$\%) & $0.70064 + 0.89846 i$ ($0.079$\%) & $0.70152 + 0.89824 i$ \\
    \hline\hline

    5& $0$ & $1.0173 + 0.0952 i$ ($0.49$\%) & $1.0123 + 0.09487 i$ ($5\cdot 10^{-6}$\%) & $1.01230 + 0.09487 i$ \\
    \hline
    & $1$ & $1.0073 + 0.2867 i$ ($0.49$\%) & $1.00222 + 0.28582 i$ ($7\cdot 10^{-6}$\%) & $1.00222 + 0.28582 i$ \\
    \hline
    & $2$ & $0.9879 + 0.4818 i$ ($0.49$\%) & $0.98269 + 0.48033 i$ ($0.0004$\%) & $0.98270 + 0.48033 i$ \\
    \hline
    & $3$ & $0.9604 + 0.6826 i$ ($0.49$\%) & $0.95496 + 0.68054 i$ ($0.004$\%) & $0.95500 + 0.68056 i$ \\
    \hline
    & $4$ & $0.9267 + 0.8908 i$ ($0.48$\%) & $0.92081 + 0.88819 i$ ($0.02$\%) & $0.92108 + 0.88820 i$ \\
    \hline\hline
    6& $0$ & $1.2162 + 0.0955 i$ ($0.35$\%) & $1.21201 + 0.09526 i$ ($2\cdot 10^{-6}$\%) & $1.21201 + 0.09527 i$ \\
    \hline
    & $1$ & $1.2078 + 0.2873 i$ ($0.35$\%) & $1.20357 + 0.28665 i$ ($3\cdot 10^{-6}$\%) & $1.20357 + 0.28665 i$ \\
    \hline
    & $2$ & $1.1914 + 0.4816 i$ ($0.35$\%) & $1.18707 + 0.48056 i$ ($0.0001$\%) & $1.18707 + 0.48056 i$ \\
    \hline
    & $3$ & $1.1677 + 0.6801 i$ ($0.35$\%) & $1.16326 + 0.67858 i$ ($0.001$\%) & $1.16327 + 0.67859 i$ \\
    \hline
    & $4$ & $1.1379 + 0.8840 i$ ($0.35$\%) & $1.13323 + 0.88207 i$ ($0.007$\%)& $1.13332 + 0.88210 i$ \\
    \hline\hline
    7& $0$ & $1.4134 + 0.0957 i$ ($0.26$\%) & $1.40974 + 0.09551 i$ ($6\cdot 10^{-7}$\%) & $1.40974 + 0.09551 i$ \\
    \hline
    & $1$ & $1.4061 + 0.2876 i$ ($0.26$\%) & $1.40247 + 0.28716 i$ ($10^{-6}$\%) & $1.40247 + 0.28716 i$ \\
    \hline
    & $2$ & $1.3919 + 0.4815 i$ ($0.26$\%) & $1.38818 + 0.48071 i$ ($0.00004$\%) & $1.38818 + 0.48071 i$ \\
    \hline
    & $3$ & $1.3711 + 0.6785 i$ ($0.26$\%) & $1.36735 + 0.67735 i$ ($0.0005$\%) & $1.36736 + 0.67735 i$ \\
    \hline
    & $4$ & $1.3446 + 0.8796 i$ ($0.26$\%) & $1.34070 + 0.87816 i$ ($0.003$\%) & $1.34074 + 0.87818 i$ \\
    \hline
\end{tabular}
\caption{The quasi-normal frequencies of a Schwarzschild black hole in units where $GM=1$, comparing our uniform approximation method, $6^{th}$ order WKB and numerical results. In parenthesis the relative error on the absolute value with respect to the numerical result $|\frac{\bar \omega_{\text{exact}}-\bar \omega_{\text{approx.}}}{\bar \omega_{\text{exact}}}|$. Note that, while we present truncated values of $\bar \omega$, we used more accurate values to compute errors.\\
The data in the second and third column of~\cref{table:linear_frequencies} (for $\bar \ell\le4$) can be found for example in~\cite{Matyjasek_2017}. We also computed the exact frequencies using the convenient implementation of Leaver's method that is found in the Black Hole Perturbation Toolkit~\cite{BHPToolkit}. Similarly,~\cite{Konoplya_2019} provides code to compute the WKB prediction. Lastly, numerical frequencies for $\bar\ell=2,\,n\le1000$ can be found in~\cite{n1000Cardoso}.}
\label{table:linear_frequencies}
\end{table}

\section{Nonlinear modes} \label{sec:nonlin}

Having proven the validity of our method to compute the linear spectrum of QNM, we now use the technique to compute the amplitude of second-order modes using the nonlinear RW/Zerilli equation described in~\cref{sec:nonlinDefs}. We first explain how to compute the nonlinear amplitude using uniform expansions before discussing the implications of our results.

\subsection{Ratio of amplitudes}

We now consider the differential equation~\eqref{eq:RWZ} for a nonlinear mode, sourced by two linear modes $\bar \Psi_1$ and $\bar \Psi_2$ of frequencies $\bar \omega_1, \bar \omega_2$ and mode numbers $(\bar \ell_1, \bar m_1, n_1)$, $(\bar \ell_2, \bar m_2, n_2)$. We also denote by $\mathcal{A}_1$ and $\mathcal{A}_2$ the amplitudes of the linear modes entering~\cref{eq:barpsi}.
Since the source term is quadratic in the linear modes we know that the frequency $\omega$ of the nonlinear mode is given by~\cite{Lagos_2023}
\begin{equation}
    \omega = \bar \omega_1 + \bar \omega_2 \; , \quad \mathrm{or} \;  \omega = \bar \omega_1 - (\bar \omega_2)^* \; ,
\end{equation}
where the star denotes complex conjugation. In the following, we will focus on the first case $\omega = \bar \omega_1+\bar \omega_2$, the treatment of the second one being very similar~\footnote{To recover the second possibility $\omega = \bar \omega_1 - (\bar \omega_2)^*$ we would have to consider a source term of the form $S = \hat S \bar \Psi_1 (\bar \Psi_2)^*$ in~\cref{eq:RWZ}.}.
On the other hand, a whole range of values of $\ell$ are allowed according to the usual rules of multiplication of spherical harmonics~\cite{Brizuela:2006ne, Brizuela:2007zza,Brizuela:2009qd, Lagos_2023}:
\begin{equation}
    | \bar \ell_1 - \bar \ell_2 | \leq \ell \leq \bar \ell_1 + \bar \ell_2 \; , \quad m = m_1+m_2 \; .
\end{equation}

We want to solve~\cref{eq:RWZ} using the same method of uniform approximation as before, by comparing it with the simpler~\cref{eq:DEphi}.
Once $\omega$ and $\ell$ are given, we can compute the value of $\nu$ for nonlinear modes following~\cref{eq:arealaw}. We give in Table~\ref{table:nonlinear_ratio} a sample of values of $\nu$ for a range of values of $\bar \ell_1$, $\bar \ell_2$ and $\ell$.
Then, using the method of variation of constants we first find the solution to the differential equation on $\phi$,~\cref{eq:DEphi}:
\begin{align} \label{eq:phinonlin}
    \frac{\phi(\sigma)}{\mathcal{A}_1 \mathcal{A}_2} &= e^{i \pi (\nu + 1/2)/2} \big[ (c_A - F_B(\sigma)) \phi_A(\sigma) + (c_B + F_A(\sigma)) \phi_B(\sigma) \big] \; , \\
    F_A(\sigma) &=  \frac{1}{\mathcal{A}_1 \mathcal{A}_2} \int_{\sigma^+}^\sigma \Lambda(\tilde \sigma) \phi_A(\tilde \sigma) \mathrm{d}\tilde \sigma \; , \quad F_B(\sigma) = \frac{1}{\mathcal{A}_1 \mathcal{A}_2} \int_{\sigma^+}^\sigma \Lambda(\tilde \sigma) \phi_B(\tilde \sigma) \mathrm{d}\tilde \sigma \; , \label{eq:FAFB}
\end{align}
where $\phi_{A/B}$ are given in~\cref{eq:phi1phi2}, and $c_{A/B}$ are two constants that we will tune in order to ensure QNM boundary conditions. Notice that we have conveniently normalized $\phi$ by the product of amplitudes $\mathcal{A}_1 \mathcal{A}_2$ to which it should be proportional.
In~\cref{app:CVint} we show that the two integrals defining $F_A$ and $F_B$ are convergent for $r_* \rightarrow \pm \infty$\footnote{More precisely, they are composed of both a convergent and divergent piece. However, the divergent part drops off in the asymptotic field $\phi$ in~\cref{eq:phinonlin}, while the convergent piece amount to sending $\sigma \rightarrow \pm \infty$, see~\cref{app:CVint} for more details. }. As we anticipated in~\cref{sec:nonlinDefs}, this is intimately related to the asymptotic features of the source term $\hat S$, and the integrals may have failed to converge had we worked in another gauge where $\hat S$ did not display these asymptotics.
Thus, it is now trivial to read the asymptotic behavior of the nonlinear mode $\Psi$ given in~\cref{eq:Psisol}, as it just parallels the case of linear modes:
\begin{align}
    \frac{\Psi}{\mathcal{A}_1 \mathcal{A}_2} &\sim \frac{e^{i \pi (\nu + 1/2)/2}}{\sqrt{2 \omega}} \bigg[ e^{-i \omega r_* -i C_\infty} \bigg( \big( c_A - F_B(\infty) \big) e^{i \pi \nu/4} + \big( c_B + F_A(\infty) \big) \frac{\sqrt{2\pi}}{\Gamma(1+ \nu)} e^{- i \pi \nu/4} \bigg) \nonumber \\
    &+ \big( c_B + F_A(\infty) \big) \, e^{i \omega r_* +i C_\infty} e^{- 3i \pi (1+ \nu)/4} \bigg] \; , \quad \quad \mathrm{for} \; r_* \rightarrow \infty \; , \\
    \frac{\Psi}{\mathcal{A}_1 \mathcal{A}_2} &\sim \frac{e^{i \pi (\nu + 1/2)/2}}{\sqrt{2 \omega}} \bigg[ e^{-i \omega r_* +i C_H} \bigg( \big( c_B + F_A(-\infty) \big) e^{i \pi (1+ \nu)/4} + \big( c_A - F_B(-\infty) \big) \frac{\sqrt{2\pi}}{\Gamma(- \nu)} e^{- i \pi (1+\nu)/4} \bigg) \nonumber \\
    &+ \big( c_A - F_B(-\infty) \big) \, e^{i \omega r_* -i C_H} e^{- 3i \pi \nu/4} \bigg] \; , \quad \quad \mathrm{for} \; r_* \rightarrow - \infty \; ,
\end{align}
where $C_\infty$ and $C_H$ can be obtained from the linear expressions in~\cref{eq:barCinfty,eq:barCH} just by replacing $\bar \omega \rightarrow \omega$,  $\bar \nu \rightarrow \nu$, $\bar \ell \rightarrow \ell$ and $\bar r_*^+ \rightarrow r_*^+$. QNM boundary conditions impose
\begin{equation}
    c_B = - F_A(\infty) \; , \quad c_A = F_B(-\infty) + i \frac{\Gamma(-\nu)}{\sqrt{2\pi}} e^{i \pi \nu/2} \big( F_A(\infty) - F_A(-\infty) \big) \; .
\end{equation}
Finally, we can get the amplitude of the nonlinear mode normalized by the product of linear modes at infinity, which is the important quantity for ringdown models in gravitational-wave observations:
\begin{align} \label{eq:ratioNumerical}
    \frac{\Psi}{\bar \Psi_1 \bar \Psi_2} &\sim \sqrt{\frac{2 \bar \omega_1 \bar \omega_2}{\bar \omega_1 + \bar \omega_2}}  e^{i( \bar C_{\infty,1} + \bar C_{\infty,2}  - C_\infty)} e^{i \pi(1+3\nu -n_1 -n_2)/4} \nonumber \\
    &\times \int_{-\infty}^\infty \frac{\Lambda(\tilde \sigma)}{\mathcal{A}_1 \mathcal{A}_2} \bigg( i \frac{\Gamma(-\nu)}{\sqrt{2\pi}} e^{i \pi \nu/2}  \phi_A(\tilde \sigma) -  \phi_B(\tilde \sigma) \bigg) \mathrm{d}\tilde \sigma \; , \quad \mathrm{for} \; r_* \rightarrow \infty \; .
\end{align}
From this equation one can numerically compute this ratio of amplitudes. Indeed, notice that the source term $\Lambda$ is related to the original source $S$ present in the RW/Zerilli equation by~\cref{eq:defLambda}, and $S$ itself depends on $r$ which is related to $\sigma$ via~\cref{eq:defsigma}. Moreover, notice that the $\sqrt{\bar \omega}$ factor is needed in order to make the ratio of amplitude dimensionless since $\Lambda$, $\mathcal{A}_1$ and $\mathcal{A}_2$ are dimensionful.

\subsection{Discussion} \label{sec:discussion}

In~\cref{table:nonlinear_ratio,table:nonlinear_ratio2} we give the numerical value of the ratio in~\cref{eq:ratioNumerical} for the Zerilli equation (i.e. both $\Psi$, $\bar \Psi_1$ and $\bar \Psi_2$ are of even parity) with a range of values of $\ell$, $\bar \ell_1$ and $\bar \ell_2$. Some regularities emerge from the table; for example, notice that while the real part of $\nu$ is nearly constant, its imaginary part is approximately equal to $\bar \ell_1 + \bar \ell_2 - \ell$. This behavior could be confirmed e.g. by looking at the eikonal limit of our computations, which we plan to investigate in a near future.
A clear trend that we can deduce from~\cref{table:nonlinear_ratio2}  is that the ratio~\eqref{eq:ratioNumerical} is maximized when the Clebsch-Gordan upper bound is saturated, i.e. $\ell = \bar \ell_1 + \bar \ell_2$~\footnote{The same conclusion cannot be inferred directly from~\cref{table:nonlinear_ratio} because the range of $\ell$ considered never satisfies this equality.}. Since the damping times of the nonlinear modes, equal to the inverse of the imaginary part of $\omega$, are all of the same order in~\cref{table:nonlinear_ratio}, this means that for fixed linear amplitudes the nonlinear modes with $\ell = \bar \ell_1 + \bar \ell_2$ should be the dominant ones.
This is a robust conclusion of our toy-model; of course, it remains to be elucidated whether this feature persists in a more realistic ringdown model of nonlinear modes, where one would have to take into account the exact expression of the source term (instead of the simple model~\cref{eq:toyModelS}) and Clebsch-Gordan coefficients. Moreover, the actual amplitude of the nonlinear modes will also depend on the product of amplitudes of linear modes $\mathcal{A}_1 \times \mathcal{A}_2$, which tends to be suppressed as one moves away from the dominant mode $\bar \ell_1, \bar \ell_2=2$ produced in black-hole mergers, at least for
approximately equal mass, quasi–circular binaries~\cite{1995PhRvD..52.2044A, 1978ApJ...225..687D,Berti:2007fi,Baibhav:2018rfk,Hughes:2019zmt}. 

Nonetheless, our results show that it should be technically possible to deduce the amplitude of nonlinear modes in a ringdown signal solely from the measurement of the linear amplitudes $\mathcal{A}_1$, $\mathcal{A}_2$. As such, we can envision two interesting applications of our method, which we plan to explore in the near future:
\begin{itemize}
    \item \textbf{Improvement of ringdown models}: Typical ringdown models depend on the $N$ complex amplitudes of the $N$ linear modes that one includes in the model~\cite{Berti:2005ys}. Within the regime of perturbation theory, it should be possible to add to the signal, on top of the linear modes, the associated nonlinear modes \textit{without introducing any free parameter}. Our approach, were the correct source $S$ in~\cref{eq:RWZ} known, would enable one to compute the amplitude of the nonlinear modes and to improve ringdown models without any additional cost from a data analysis perspective.
    \item \textbf{Tests of General Relativity}: Alternatively, one can view our results as hinting properties of the nonlinear-to-linear amplitude ratio in GR. For instance, it should be possible to include additional free amplitudes in a ringdown model at the frequencies of nonlinear modes, in the spirit of~\cite{Mitman:2022qdl, Cheung:2022rbm}, and compare the value obtained from data to the GR prediction. This would constitute a new test probing GR further into its non-linear regime. 

\end{itemize}

\begin{table}[htbp]
\centering
\begin{tabular}{ccc|c|c|c}
	\hline
	$\ell$ & $\bar \ell_1$ & $\bar \ell_2$ & $\omega$ & $\nu$ & $\Psi/ (\bar \Psi_1 \bar \Psi_2)$  \\
	\hline
    \hline
    2 & 2 & 2 & $0.771+0.182i$ & $0.337-1.92i$ & $(5.39-1.19i)\times 10^{-2}$ \\
     & & 3 & $0.993+0.184i$ & $0.314 - 2.91i$ & $(3.35-0.39i)\times 10^{-2}$ \\
     & & 4 & $1.20+0.186i$ & $0.298-3.81i$ & $(2.34-0.20i)\times 10^{-2}$ \\
     \hline
     & 3 & 3 & $1.21+0.187i$ & $0.304-3.88i$ & $(2.32-0.18i)\times 10^{-2}$ \\
     & & 4 & $1.42+0.188i$ & $0.295-4.76i$ & $(1.72-0.12i)\times 10^{-2}$ \\
     & & 5 & $1.62+0.189i$ & $0.288-5.61i$ & $(1.33-0.10i)\times 10^{-2}$ \\
     \hline
     & 4 & 4 & $1.63+0.189i$ & $0.290-5.63i$ & $(1.33-0.10i)\times 10^{-2}$ \\
     & & 5 & $1.83+0.190i$ & $0.287-6.47i$ & $(1.06-0.09i)\times 10^{-2}$ \\
     & & 6 & $2.03+0.190i$ & $0.283-7.29i$ & $(8.71-0.87i)\times 10^{-3}$ \\
    \hline
    \hline
     3 & 2 & 2 &  $0.771+0.182i$ & $0.409-0.860i$ & $(7.12-3.68i)\times 10^{-2}$ \\
     & & 3 & $0.993+0.184i$ & $0.375-1.93i$ & $(4.16-0.73i)\times 10^{-2}$ \\
     & & 4 & $1.20+0.186i$ & $0.352-2.90i$ & $(2.71-0.30i)\times 10^{-2}$ \\
     & & 5 & $1.40+0.186i$ & $0.336-3.81i$ & $(1.93-0.17i)\times 10^{-2}$ \\
     \hline
     & 3 & 3 & $1.21+0.187i$ & $0.357-2.96i$ & $(2.68-0.25i)\times 10^{-2}$ \\
     & & 4 & $1.42+0.188i$ & $0.343-3.90i$ & $(1.91-0.15i)\times 10^{-2}$ \\
     & & 5 & $1.62+0.189i$ & $0.332-4.80i$ & $(1.45-0.11i)\times 10^{-2}$ \\
     & & 6 & $1.83+0.189i$ & $0.323-5.67i$ & $(1.14-0.09i)\times 10^{-2}$ \\
     \hline
     & 4 & 4 & $1.63+0.189i$ & $0.334-4.83i$ & $(1.44-0.11i)\times 10^{-2}$ \\
     & & 5 & $1.83+0.190i$ & $0.326-5.71i$ & $(1.14-0.09i)\times 10^{-2}$ \\
     & & 6 & $2.03+0.190i$ & $0.319-6.57i$ & $(9.19-0.85i)\times 10^{-3}$ \\
     & & 7 & $2.29+0.190i$ & $0.314-7.42i$ & $(7.57-0.83i)\times 10^{-3}$ \\
    \hline
\end{tabular}
\caption{Values of $\omega$, $\nu$ and of the ratio of nonlinear to linear amplitudes $\Psi/ (\bar \Psi_1 \bar \Psi_2)$ when $\ell=2,3$ and both $\Psi$, $\bar \Psi_1$ and $\bar \Psi_2$ are of even parity, in units where $GM=1$. We only give the values where the Clebsch-Gordan coefficient is nonzero i.e. $| \bar \ell_1 - \bar \ell_2 | \leq \ell \leq \bar \ell_1 + \bar \ell_2$, and we restrict to observationally relevant modes with $\ell, \bar \ell_1, \bar \ell_2 \geq 2$. Moreover, we consider only second-order QNMs sourced by fundamental modes, i.e. $n_1=n_2=0$.}
\label{table:nonlinear_ratio}
\end{table}
\begin{table}[htbp]
\centering
    \begin{tabular}{ccc|c|c|c}
    \hline
	$\ell$ & $\bar \ell_1$ & $\bar \ell_2$ & $\omega$ & $\nu$ & $\Psi/ (\bar \Psi_1 \bar \Psi_2)$  \\
	\hline
    \hline
     4 & 2 & 2 &  $0.771+0.182i$ & $0.472+0.212i$ & {\color{blue}$(-1.73+0.01i)\times 10^{-1}$} \\
     & & 3 & $0.993+0.184i$ & $0.423-0.926i$ & $(5.54-2.16i)\times 10^{-2}$ \\
     & & 4 & $1.20+0.186i$ & $0.391-1.94i$ & $(3.41-0.54i)\times 10^{-2}$ \\
     & & 5 & $1.40+0.186i$ & $0.368-2.89i$ & $(2.27-0.24i)\times 10^{-2}$ \\
     & & 6 & $1.60+0.186i$ & $0.351-3.81i$ & $(1.64-0.14i)\times 10^{-2}$ \\
     \hline
     & 3 & 3 & $1.21+0.187i$ & $0.396-2.01i$ & $(3.35-0.45i)\times 10^{-2}$ \\
     & & 4 & $1.42+0.188i$ & $0.376-2.99i$ & $(2.24-0.20i)\times 10^{-2}$ \\
     & & 5 & $1.62+0.189i$ & $0.360-3.91i$ & $(1.63-0.13i)\times 10^{-2}$ \\
     & & 6 & $1.83+0.189i$ & $0.347-4.81i$ & $(1.24-0.10i)\times 10^{-2}$ \\
     & & 7 & $2.02+0.189i$ & $0.338-5.69i$ & $(9.90-0.94i)\times 10^{-3}$ \\
     \hline
     & 4 & 4 & $1.63+0.189i$ & $0.362-3.94i$ & $(1.63-0.13i)\times 10^{-2}$ \\
     & & 5 & $1.83+0.190i$ & $0.351-4.85i$ & $(1.24-0.10i)\times 10^{-2}$ \\
     & & 6 & $2.03+0.190i$ & $0.351-5.74i$ & $(9.88-0.93i)\times 10^{-3}$ \\
     & & 7 & $2.29+0.190i$ & $0.333-6.60i$ & $(8.07-0.85i)\times 10^{-3}$ \\
     & & 8 & $2.42+0.190i$ & $0.347-7.47i$ & $(6.72-0.82i)\times 10^{-3}$ \\
     \hline
     \hline
     5 & 2 & 3 & $0.993+0.184i$ & $0.473+0.109i$ & {\color{blue}$-(1.21+0.19i)\times 10^{-1}$} \\
     & & 4 & $1.20+0.186i$ & $0.431-0.954i$ & $(4.65-1.58i)\times 10^{-2}$ \\
     & & 5 & $1.40+0.186i$ & $0.402-1.95i$ & $(2.89-0.44i)\times 10^{-2}$ \\
     & & 6 & $1.60+0.186i$ & $0.389-2.90i$ & $(1.95-0.21i)\times 10^{-2}$ \\
     & & 7 & $1.80+0.187i$ & $0.363-3.82i$ & $(1.43-0.13i)\times 10^{-2}$ \\
     \hline
     & 3 & 3 & $1.21+0.187i$ & $0.436-1.02i$ & $(4.52-1.26i)\times 10^{-2}$ \\
     & & 4 & $1.42+0.188i$ & $0.409-2.04i$ & $(2.84-0.34i)\times 10^{-2}$ \\
     & & 5 & $1.62+0.189i$ & $0.389-3.00i$ & $(1.93-0.17i)\times 10^{-2}$ \\
     & & 6 & $1.83+0.189i$ & $0.373-3.93i$ & $(1.42-0.12i)\times 10^{-2}$ \\
     & & 7 & $2.02+0.189i$ & $0.360-4.83i$ & $(1.09-0.09i)\times 10^{-2}$ \\
     & & 8 & $2.22+0.189i$ & $0.349-5.72i$ & $(8.75-0.93i)\times 10^{-3}$ \\
     \hline
     & 4 & 4 & $1.63+0.189i$ & $0.391-3.03i$ & $(1.92-0.17i)\times 10^{-2}$ \\
     & & 5 & $1.83+0.190i$ & $0.376-3.97i$ & $(1.41-0.12i)\times 10^{-2}$ \\
     & & 6 & $2.03+0.190i$ & $0.363-4.88i$ & $(1.09-0.09i)\times 10^{-2}$ \\
     & & 7 & $2.29+0.190i$ & $0.353-5.77i$ & $(8.73-0.92i)\times 10^{-3}$ \\
     & & 8 & $2.42+0.190i$ & $0.344-6.64i$ & $(7.20-0.84i)\times 10^{-3}$ \\
     & & 9 & $2.62+0.191i$ & $0.337-7.50i$ & $(6.00-0.83i)\times 10^{-3}$ \\
     \hline
     \hline
\end{tabular}
\caption{Same as~\cref{table:nonlinear_ratio}, but for $\ell=4,5$. We have highlighted in blue the values of the ratio whose order-of-magnitude is $10^{-1}$, which precisely correspond to the values for which $\ell = \ell_1+\ell_2$. At the moment we do not have any physical explanation on the sign difference of the blue ratios, which seems to be related to a slower decay of the integrand when $\ell = \ell_1+\ell_2$.}
\label{table:nonlinear_ratio2}
\end{table}

\section{Accuracy of uniform expansions} \label{sec:accuracy}

\begin{figure}
    \centering
    \includegraphics[width=0.5\linewidth]{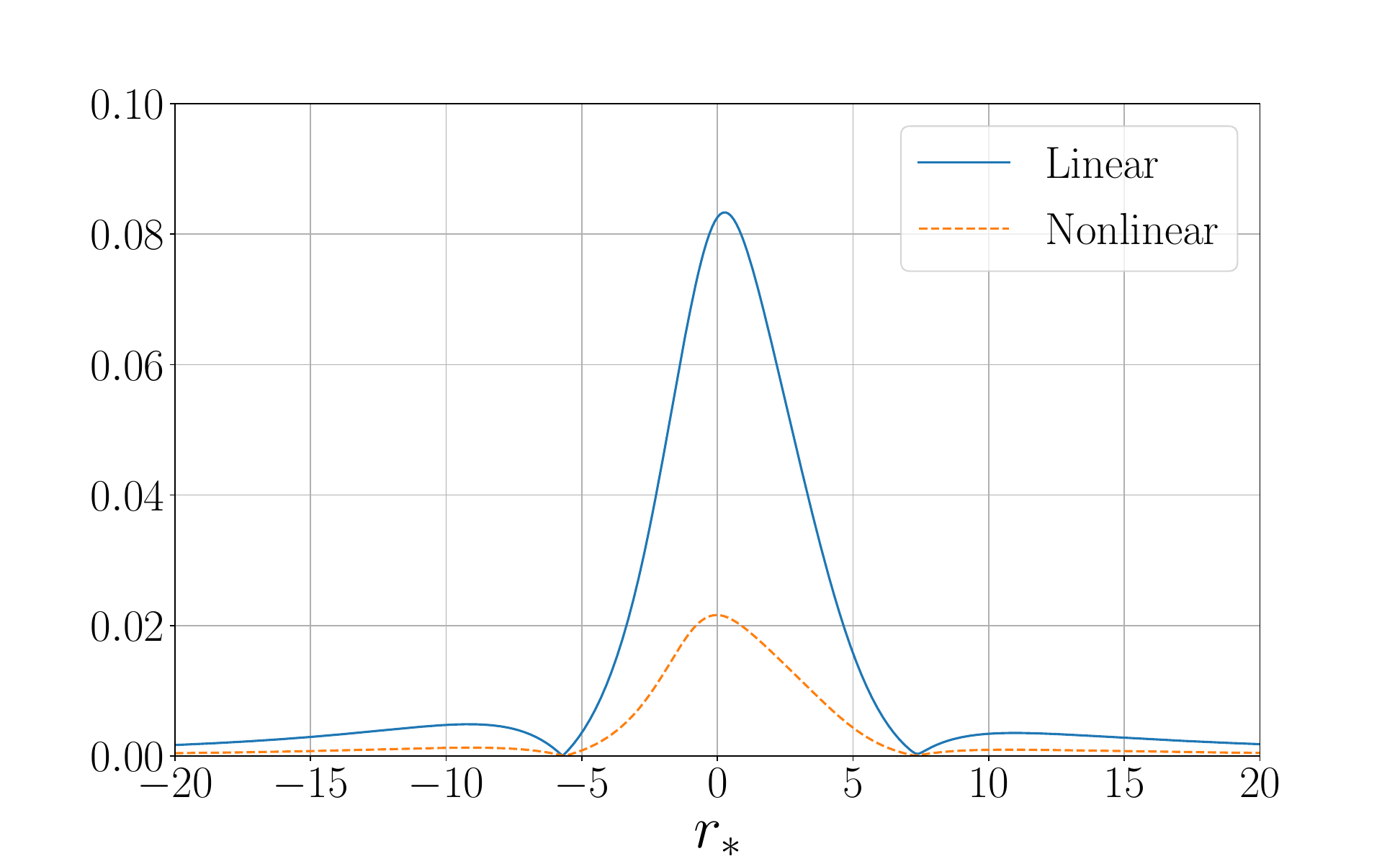}
    \caption{\textit{Accuracy of the uniform approximation}: We plot the ratio~\eqref{eq:validityApprox}  both for the $\bar \ell=2, n=0$ linear solution and the $\ell=4, \bar \ell_1=\bar \ell_2=2$, $n_1=n_2=0$ nonlinear solution, as a function of $r_*$ in units where $GM=1$. The ratio is always small, signalling that the error on QNMs frequencies and nonlinear amplitudes is at the percent level.  }
    \label{fig:accuracy}
\end{figure}

All our previous results have been obtained using the method of uniform expansions which is itself an approximation. It is thus quite natural to ask what is the accuracy of this approximation. 
Going back to~\cref{sec:uniform}, we see that the \textit{only} approximation we have made so far is to neglect the second term on the right-hand side of~\cref{eq:fullEqSigma} in order to be able to compute $\sigma'$. This is valid when
\begin{equation} \label{eq:validityApprox}
    \Bigg| \frac{1}{2} \frac{(\sigma')^{1/2}}{\omega^2 - V} \frac{\mathrm{d}^2}{\mathrm{d}r_*^2} (\sigma')^{-1/2} \Bigg| \ll 1 \; .
\end{equation}
Once we have obtained the numerical solution for $\sigma(r)$ from~\cref{eq:defsigma}, it is easy to check this inequality. In~\cref{fig:accuracy} we show the magnitude of the ratio in~\cref{eq:validityApprox} both for the $\bar \ell=2, n=0$ linear solution and the $\ell=4, \bar \ell_1=\bar \ell_2=2$, $n_1=n_2=0$ nonlinear solution, as a function of $r_*$. In the linear case, the ratio~\eqref{eq:validityApprox} reaches a maximal value of $\simeq 0.08$. Notice, however, that it does not translate directly into a $8$\% error in the estimation of the QNM frequencies, as from~\cref{table:linear_frequencies} the error of our method, compared to a more accurate numerical result, is only of $3$\% for the linear $\bar \ell=2, n=0$ QNM frequency.

In the nonlinear case, the approximation seems to work much better, as the maximum of the ratio~\eqref{eq:validityApprox} is only of $0.02$. Since we cannot yet precisely compare our results to the numerical relativity ones in e.g.~\cite{Mitman:2022qdl, Cheung:2022rbm} ~\footnote{We would need to plug in the exact expression of the source term in the right gauge and derive the asymptotic waveform; notice that this source term generically contains even and odd-parity first-order perturbations~\cite{Brizuela:2009qd}, unlike the example discussed in~\cref{app:source}}, it remains quite difficult to estimate the actual error in the ratio of nonlinear-to-linear amplitudes,~\cref{eq:ratioNumerical}, but we can give an upper bound of $2$\% in this particular case. This accuracy could then be improved by computing next-to-leading order terms in the uniform expansion as is done in e.g.~\cite{doi:10.1143/JPSJ.14.1771,1971JChPh..54.3864P,PhysRev.106.1156}. Whether or not these higher-order corrections would be needed in ringdown models given the sensitivity of gravitational-wave detectors such as LISA~\cite{Robson:2018ifk} is an interesting question that we leave to further work.  

\section{Conclusions}

In this work we have discussed the application of the so-called method of uniform approximation to black hole perturbation theory. 
As we have shown, this method provides reliable predictions of both the linear quasi-normal frequencies~(\cref{table:linear_frequencies}), and the amplitude of the nonlinear modes~(\cref{table:nonlinear_ratio,table:nonlinear_ratio2}). Concerning the linear spectrum, one important advantage of this technique is providing accurate formulas for the QNM frequencies even at large overtone number $n$, while the WKB approach loses accuracy as $n$ increases. Other advantages of our formalism compared to WKB are that it does not require any matching between approximate solutions in different regions of the integration domain, and its relative simplicity and compact formulas. Contrary to what was done in~\cite{Hatsuda:2021gtn} where $V(r)$ was then expanded around the minimum, we preferred to numerically evaluate all integrals to maximize precision.

We then investigated the predictions of our method for the amplitude of second-order perturbations sourced by linear QNMs.
For simplicity, we assumed a toy model source~\cref{eq:toyModelS} for the RW/Zerilli equation at second order in the main body of our work. As we show in appendix~\ref{app:source} for the case of $\ell=4, \bar\ell_1=\bar\ell_2=2$, the same method can be straightforwardly applied to the full GR source, when the dynamical variables are chosen in such a way that this source displays the appropriate falloff conditions. In the context of this particular toy-model, our numerical results suggest that the amplitude of nonlinear QNMs is maximized when the right Clebsch-Gordan inequality is saturated, $\ell = \bar\ell_1+\bar\ell_2$. A realistic source will have a more complex dependence on $ \ell,\ell_1,\ell_2$, making it difficult to establish at this point whether the same happens in GR or not. 
We evaluated the accuracy of our technique in section \ref{sec:accuracy}, finding an error of at most the percent level due to the uniform approximation, with respect to an exact solution of the quadratic problem with the toy model source.

The wavefunction can also be approximately computed, both numerically at finite $r_*$ and analytically in the limit of $r_*\rightarrow\pm\infty$. We believe one of the main strengths of the method to be the analytic control that it provides on the problem, a feature that could allow one to analytically study \cref{eq:quantization} and \cref{eq:ratioNumerical} for excited overtones $n\rightarrow\infty$ or in the eikonal limit $\ell\rightarrow\infty$.

Our results indicate that the amplitude of second-order QNMs, properly normalized by the product of amplitudes of the linear modes which generate it, can be computed in General Relativity. Thus, it should be possible to use this fact in order to improve ringdown modelling or design tests of GR, as we discussed in~\cref{sec:discussion}. For this we would need to derive the expression of the source term in the RW/Zerilli-type equation in a suitable gauge, and then relate the amplitude of the RW/Zerilli-type scalar to the asymptotic waveform, a task which we plan to undertake soon. Moving to more and more realistic scenarios, it would be interesting to employ our method in the case of Kerr spacetime. Even more ambitiously, it would be interesting to model the source terms in cases of departure from GR and standard BH geometry, e.g. using the formalism in \cite{Franciolini:2018uyq} and leveraging as much as possible the properties of the near-light ring geometry.

\acknowledgments
We wish to thank D.Perrone, T. Barreira, A. Kehagias and A. Riotto for useful discussions and for sharing their draft of~\cite{perrone2023nonlinear}. We would also like to thank Macarena Lagos and Lam Hui for discussions.
This work makes use of the Black Hole Perturbation Toolkit~\cite{BHPToolkit}.  

\appendix

\section{Exact source term in a particular case} \label{app:source}

In this Appendix we will benchmark our toy-model for $\hat S$,~\cref{eq:defSHat}, by comparing it to the expression given in Ref.~\cite{Nakano:2007cj}, valid for two even first-order perturbations with $\bar \ell_1=\bar \ell_2=2$, $\bar m_1=\bar m_2=\pm 2$, $n_1=n_2$ (since the two modes are the same, we will simply denote them by $\bar \Psi$), generating an even second-order perturbation with $\ell=4$, $m=\pm 4$:
\begin{equation} \label{eq:ShatNakano}
    \hat S_\mathrm{exact} = f_1(r) + f_2(r) \frac{\bar \Psi'}{\bar \Psi} + f_3(r) \bigg( \frac{\bar \Psi'}{\bar \Psi} \bigg)^2 \; ,
\end{equation}
where a prime denotes differentiation with respect to $r$, and the functions $f_1$, $f_2$, $f_3$ are given by \small
\begin{align}
    &f_1(r) = \frac{-i \bar\omega}{126} \sqrt{\frac{70}{\pi}} \bigg[ - \bar \omega^2 \frac{r(7r+4G M)}{r - 2GM} \nonumber \\
    &+ 3 \frac{276r^7+476r^6 GM-1470r^5G^2M^2-1389r^4G^3M^3-816r^3G^4M^4-800r^2G^5M^5-555rG^6M^6-96G^7M^7}{r^3(3r+GM)^2(2r+3GM)^2(r-2GM)} \nonumber \\
    &+ 9 \frac{r-2GM}{\bar \omega^2 r^7 (3r+GM)^2(2r+3GM)^4} \big(2160r^9+11760r^8GM+30560r^7G^2M^2+41124r^6G^3M^3+31596r^5G^4M^4 \nonumber \\
    &+11630r^4G^5M^5-1296r^3G^6M^6-4182r^2G^7M^7-1341rG^8M^8-144G^9M^9 \big)\bigg] \; , \\
    &f_2(r) =  \frac{-i \bar \omega}{126} \sqrt{\frac{70}{\pi}} \bigg[ -4 G^2 M^2 \frac{(r-2GM)^2}{r^2(3r+GM)^2} + \frac{6(r-2GM)}{\bar \omega^2r^6(3r+GM)^2(2r+3GM)^3} \big( 144 r^8+4116r^7GM \nonumber\\
    &+2154r^6G^2M^2-2759r^5G^3M^3-8230r^4G^4M^4-9512r^3G^5M^5-3540r^2G^6M^6-1119rG^7M^7-144G^8M^8 \big) \bigg] \; , \\
    &f_3(r) =  \frac{-i \bar \omega}{126} \sqrt{\frac{70}{\pi}} \bigg[ - \frac{(r-2GM)(7r+4GM)}{r} - \frac{3(r-2GM)}{\bar \omega^2 r^5 (3r+GM)^2(2r+3GM)^2} \big( 228r^7+8r^6GM \nonumber\\
    &-370r^5G^2M^2+142r^4G^3M^3-384r^3G^4M^4-514r^2G^5M^5-273rG^6M^6-48G^7M^7 \big) \bigg] \; .
\end{align}
\normalsize
Notice that we have both a sign difference and a $\bar \omega^{-2}$ factor with respect to Ref.~\cite{Nakano:2007cj} because of our different Fourier convention and normalization of the fields, and that we corrected a typo in the denominator of the first function $f_1$ (we replaced the $3r-GM$ by $3r+GM$ otherwise $\hat S$ would not have the correct asymptotic limit). Notice also that the most generic source term can \textit{a priori} contain odd parity first-order perturbations~\cite{Brizuela:2009qd}, but the expression~\ref{eq:ShatNakano} given in Ref.~\cite{Nakano:2007cj} ignores these terms for simplicity. 

\begin{figure}
    \centering
    \includegraphics[width=0.75\linewidth]{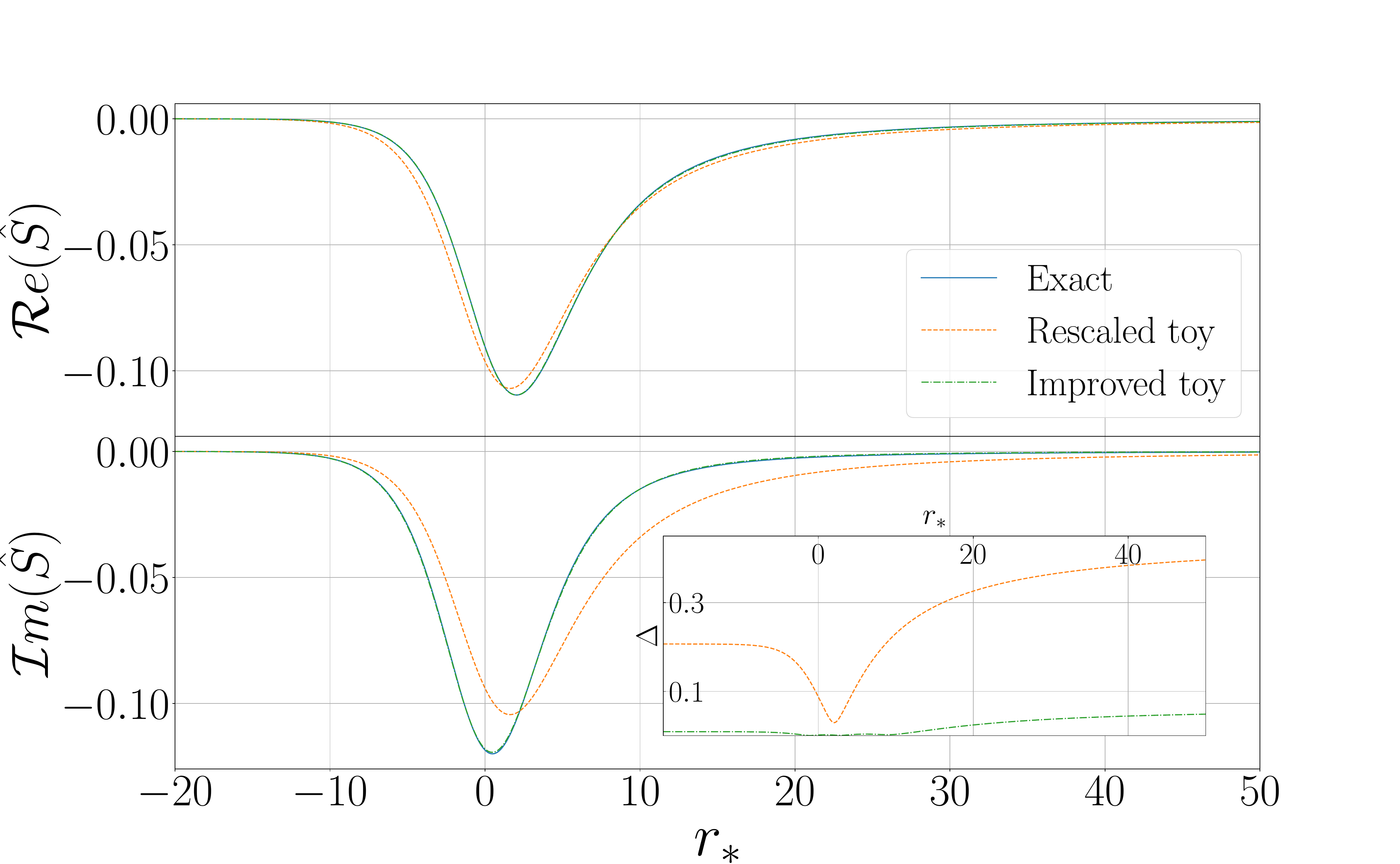}
    \caption{\textit{Exact source term}: Plot of the real and imaginary parts of the exact source term $\hat S_\mathrm{exact}$ in~\cref{eq:ShatNakano} versus a rescaled and improved version of the toy-model defined in~\cref{eq:rescaled,eq:improved}, as a function of the tortoise coordinate $r_*$ and in units in which $GM=1$.  The inset shows the fractional difference $\Delta = | (\hat S - \hat S_\mathrm{exact})/(\hat S + \hat S_\mathrm{exact}) |$. }
    \label{fig:shat}
\end{figure}

The asymptotic behavior of $\hat S$ written in~\cref{eq:asymptoticsShat} is not manifest in~\cref{eq:ShatNakano}, and is obtained only after subtle cancellations involving the asymptotic expansion of $\bar \Psi$, as explained in~\cite{Nakano:2007cj}. Indeed, notice that the asymptotic limits of the $f_i$ functions are
\begin{align}
    f_1 &\sim f_3 \sim \mathcal{O}(r) \; , \quad f_2 \sim \mathcal{O}\bigg( \frac{1}{r^2} \bigg) \; , \quad \text{for }r \rightarrow \infty \; , \\
    f_1 &\sim \mathcal{O} \bigg( \frac{1}{r-2GM} \bigg) \; , \quad f_2 \sim f_3 \sim \mathcal{O}(r-2GM) \; , \quad \text{for } r \rightarrow 2GM \; .
\end{align}
The precision to which $\bar \Psi$ is needed in order to ensure this cancellation is challenging for our numerical solution for $\bar \Psi$ written in~\cref{eq:barpsi}. However, there are other techniques such as Leaver's method~\cite{leaver} which are very efficient for getting the numerical profile of linear quasi-normal modes. We have thus chosen to implement Leaver's algorithm as presented in e.g.~\cite{leaver, Nakano:2007cj} in order to get the numerical profile $\bar \Psi$. Results are shown in~\cref{fig:shat}, where we show the real and imaginary parts of our numerical solution for $\hat S$. We also plot a <<rescaled>> version of our toy model, defined by
\begin{equation} \label{eq:rescaled}
    \hat S_\mathrm{rescaled}(r) =  \frac{a}{r^2} \bigg( 1 - \frac{2GM}{r} \bigg) \; , \quad a \simeq -2.7 - 2.8i \; ,
\end{equation}
and an <<improved>> version of the toy model,
\begin{align} \label{eq:improved}
     \hat S_\mathrm{improved}(r) &=  \frac{1}{r^2} \bigg( 1 - \frac{2GM}{r} \bigg)  \bigg( b_0 + b_1 \frac{GM}{r} + b_2 \frac{G^2 M^2}{r^2} \bigg) \; , \nonumber \\
     b_0 &\simeq -2.0 - 0.2i, \; b_1 \simeq -8.3 -8.0i , \; b_2 \simeq 16 - 1.5i \; ,
\end{align}
where in both cases the fitting constants are obtained by a least-square algorithm. As can be seen, even the simple rescaled version of the toy model is quite good at approximating the true source term, while the improved version is essentially identical.

Finally, we have also applied our method described in~\cref{sec:nonlin} in order to get the ratio of nonlinearity with the different profiles for $\hat S$:
\begin{equation}
    \frac{\Psi}{\bar \Psi^2} \Bigg|_{\hat S_\mathrm{exact}} \simeq - 0.091 + 0.463i \; , \quad \frac{\Psi}{\bar \Psi^2} \Bigg|_{\hat S_\mathrm{rescaled}} \simeq 0.477+0.488i \; , \quad \frac{\Psi}{\bar \Psi^2} \Bigg|_{\hat S_\mathrm{improved}} \simeq - 0.093 + 0.462i \; ,
\end{equation}
The absolute value $\sim 0.472$ of our result with the exact profile for $\hat S$ matches the estimates found in e.g.~\cite{Nakano:2007cj,perrone2023nonlinear} with a $\sim 10$\% discrepancy . Because of the oscillatory nature of the integral involved in~\cref{sec:nonlin}, the result for rescaled toy model is quite off the exact prediction, however the improved toy model fully recovers the exact result.

\section{Convergence of integrals} \label{app:CVint}

In this appendix we will show that the two integrals defining $F_A$ and $F_B$,~\cref{eq:FAFB}, are composed of a divergent piece and a convergent integral when the tortoise coordinate is large, $r_* \rightarrow \pm \infty$. However,  it's important to note that the divergent part just drops off in the asymptotic limit of the field $\phi$ in~\cref{eq:phinonlin}, leaving us with a finite value.

To see this, let us first note that the coordinate $\sigma$ is not real when $|r_*|$ is large, since we have the relations~\cref{eq:barsigmainfty,eq:barsigmaminfty} and $\omega$ is complex. Let us first focus on the limit $r_* \rightarrow \infty$, the treatment of the limit $r_* \rightarrow -\infty$ being similar.  When sending $r_*$ to real infinity, the coordinate $\sigma$ does instead go to a complex infinity following $\sigma \simeq 2 (\omega r_*)^{1/2}$. Because of the imaginary part of $\omega$, the integral defining $F_{A/B}$ is in fact exponentially divergent. Let us now show that this divergence is immaterial and that we could as well assume that $\sigma \rightarrow \infty$. 

Let us introduce a \textit{real} coordinate $\sigma^R$ \textquotedblleft close\textquotedblright{} to $\sigma$ at large $r_*$. We choose $\sigma^R = 2 (|\omega| r_*)^{1/2}$ for definiteness, but we could have worked with other choices as well. We now split the integral defining $F_{A/B}$ in two pieces, $\int_{\sigma^+}^\sigma = \int_{\sigma^+}^{\sigma^R} + \int_{\sigma^R}^\sigma$ (assuming analyticity of the integrand in the region of interest). Our aim is to first show that the second piece $J_{A/B}=\int_{\sigma^R}^\sigma$ gives a subleading contribution to $\phi(\sigma)$,~\cref{eq:phinonlin}, in the limit $r_* \rightarrow \infty$. Noting that both $\sigma^R$ and $\sigma$ are large, we can take the large-$r_*$ limit in the integrand to find:
\begin{equation}
    J_A \simeq \mathrm{Cst} \times \int_{r_* |\omega|/\omega}^{r_*} \frac{e^{-2 i \omega \tilde r_*}}{\tilde r_*^2} \mathrm{d}\tilde r_*
\end{equation}
where we have changed the variable of integration from $\tilde \sigma$ to $\tilde r_*$, used the asymptotic limit $\hat S \simeq r_*^{-2}$ for $r_* \rightarrow \infty$ and from now on in this appendix we will not compute the exact value of the constants in front of the dominant term of the asymptotic expansions.
The integral can be performed exactly. The lower bound vanishes for large $r_*$, while the upper bound still diverges exponentially because of the imaginary part of $\omega$. Similarly, we find
\begin{equation}
    J_B \simeq e^{-i \pi \nu/2} \frac{\sqrt{2\pi}}{\Gamma(1+\nu)} J_A +  \mathrm{Cst} \times \int_{r_* |\omega|/\omega}^{r_*} \frac{1}{\tilde r_*^2} \mathrm{d}\tilde r_*
\end{equation}
This time the second integral completely vanishes at both ends. Plugging this result into~\cref{eq:phinonlin} defining $\phi$, we get
\begin{equation}
    J_A \phi_B - J_B \phi_A \simeq \mathcal{O}\big( r_*^{-9/4}\big) \times e^{-i \omega r_*}
\end{equation}
This shows that, at large distances, the contribution to $\Psi$ of the second integral, from $\sigma_R$ to $\sigma$,  is (from \cref{eq:Psisol}) $\mathcal{O}\big( r_*^{-2}\big) \times e^{-i \omega r_*}$, i.e. an outgoing wave suppressed by the falloff of the source term. This is asymptotically negligible compared to the standard outgoing wave $e^{-i \omega r_*}$ and does not change the ratio of amplitudes at large distances. The reasoning is exactly the same for the limit $r_* \rightarrow - \infty$. Notice that the falloff conditions assumed by the source term are quite crucial in order for these terms not to contribute.

\begin{figure}
    \centering
    \includegraphics[width=0.5\linewidth]{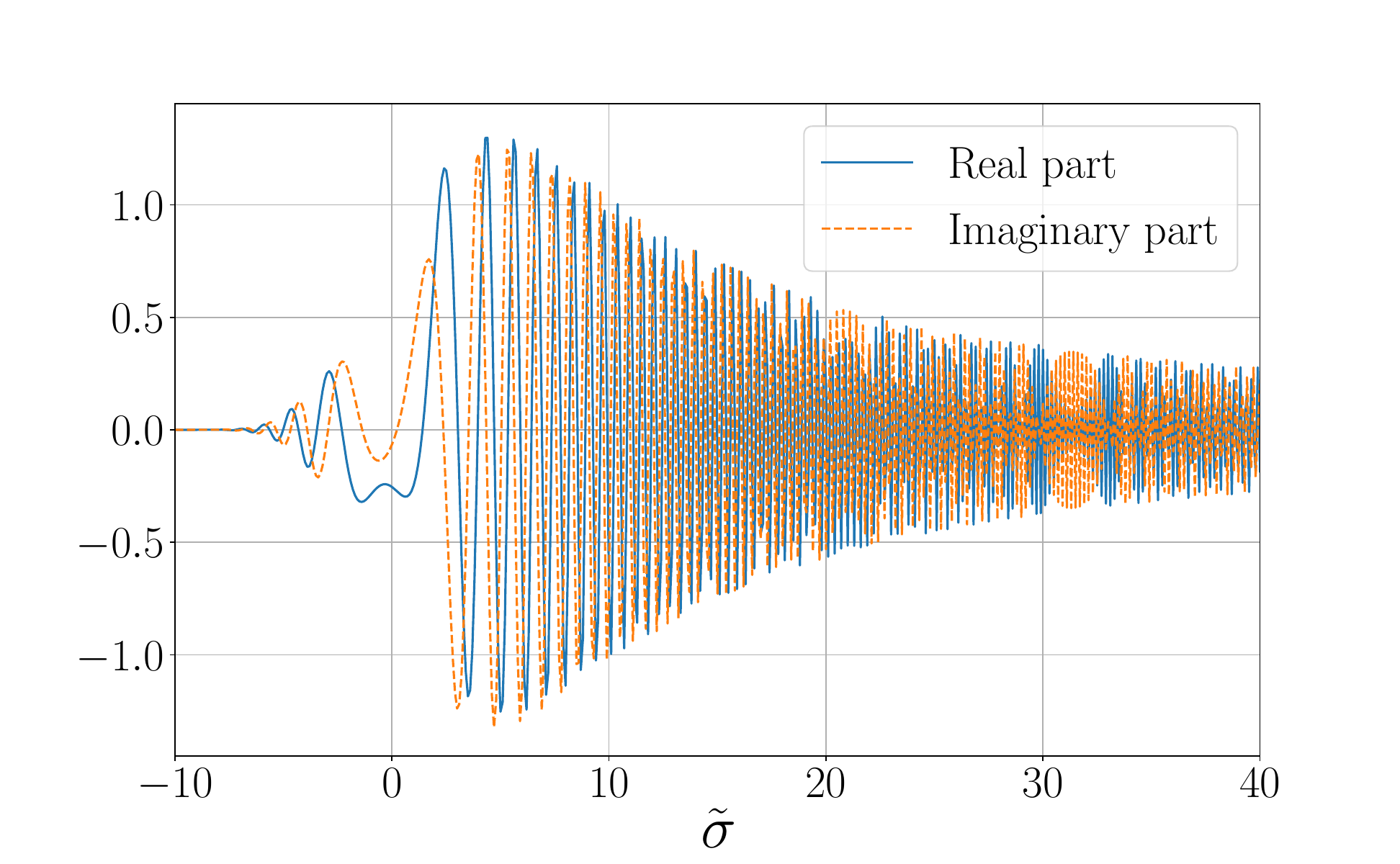}
    \caption{Plot of the real and imaginary parts of the integrand in~\cref{eq:ratioNumerical} in the particular case $\ell=4, \bar \ell_1=\bar \ell_2=2$, $\bar n_1=\bar n_2=0$, as a function of the coordinate of integration $\tilde \sigma$, in units where $GM=1$. }
    \label{fig:integrand}
\end{figure}

Let us now show that the first piece of the integral defining $F_{A/B}$ ,~\cref{eq:FAFB} with $\sigma^R$ as upper bound, is convergent for $\sigma^R \rightarrow \pm \infty$. To avoid clutter, we will from now on remove the $R$ superscript on $\sigma$, keeping in mind that we are sending $\sigma$ to \textit{real} infinity.
Let us first evaluate the convergence as $\sigma \rightarrow \infty$. We have to relate the asymptotic behavior of the background coordinates $\bar \sigma_1$ and $\bar \sigma_2$ to $\sigma$; this can be easily obtained from~\cref{eq:barsigma2infty}. Using~\cref{eq:barpsi,eq:Dnuinfty} we get
\begin{equation}
    \frac{\bar \Psi_1 \bar \Psi_2}{\mathcal{A}_1 \mathcal{A}_2} \sim \mathrm{Cst} \times \sigma^{\nu +1/2} e^{-i \sigma^2/4} \; ,
\end{equation}
Moreover, from~\cref{eq:asymptoticsShat} we know that $\hat S = \mathcal{O}(r^{-2}) = \mathcal{O}(\sigma^{-4})$ for large $\sigma$. Finally combining the definition of $\Lambda = S / (\sigma ')^{3/2}$ and the factorization $S = \hat S \Psi_1 \Psi_2$, we get to
\begin{equation}
   \frac{ \Lambda(\sigma)}{\mathcal{A}_1 \mathcal{A}_2} \sim \mathrm{Cst} \times \sigma^{\nu -2} e^{-i \sigma^2/4} \; , \quad \mathrm{for} \; \sigma \rightarrow \infty \; .
\end{equation}
 Thus,
\begin{align}
     \frac{ \Lambda(\sigma)}{\mathcal{A}_1 \mathcal{A}_2} \phi_A(\sigma) &\sim \mathrm{Cst}   \times \sigma^{2\nu -2} e^{-i \sigma^2/2} \; ,\\
     \frac{ \Lambda(\sigma)}{\mathcal{A}_1 \mathcal{A}_2} \phi_B(\sigma) &\sim \mathrm{Cst}   \times \sigma^{2\nu -2} e^{-i \sigma^2/2} + \mathrm{Cst} \times  \sigma^{-3} \; .
\end{align}
This means that the integral defining $F_A$ and $F_B$ converges as long as $\mathrm{Re}(\nu) < 3/2$, which is verified for all the values of $\nu$ that we consider. Now for $\sigma \rightarrow - \infty$ we get
\begin{equation}
    \frac{\bar \Psi_1 \bar \Psi_2}{\mathcal{A}_1 \mathcal{A}_2} \sim \mathrm{Cst} \times |\sigma|^{\nu +1/2} e^{-i \sigma^2/4} \; .
\end{equation}
Since now $\hat S = \mathcal{O} (1-2GM/r) = \mathcal{O}\big(e^{-\sigma^2/(8GM\omega)} \big)$ for $\sigma \rightarrow - \infty$ (with $\mathrm{Re} \, \omega >0$), we get much better convergence as $\sigma \rightarrow - \infty$:
\begin{align}
     \frac{ \Lambda(\sigma)}{\mathcal{A}_1 \mathcal{A}_2} \phi_A(\sigma) &\sim \mathrm{Cst}   \times |\sigma|^{2\nu +2} e^{-i \sigma^2/2} e^{-\sigma^2/(8GM\omega)} + \mathrm{Cst} \times  \sigma e^{-\sigma^2/(8GM\omega) } \; , \\
     \frac{ \Lambda(\sigma)}{\mathcal{A}_1 \mathcal{A}_2} \phi_B(\sigma) &\sim \mathrm{Cst} \times  \sigma e^{-\sigma^2/(8GM\omega)} \; ,
\end{align}
so that the integral converges whatever the value of $\nu$ is. In~\cref{fig:integrand} we show a plot of the integrand in~\cref{eq:ratioNumerical} in the particular case $\ell=4, \bar \ell_1=\bar \ell_2=2$, $\bar n_1=\bar n_2=0$. It clearly shows the asymptotic behavior derived analytically in this appendix.

\bibliographystyle{utphys}
\bibliography{refs}

\end{document}